\documentclass[a4paper,9pt]{article}
\usepackage[utf8]{inputenc}
\usepackage{url}
\usepackage{algorithm}
\usepackage{algorithmicx}
\usepackage{algpseudocode}
\usepackage{graphics}
\usepackage{graphicx}
\usepackage{color}
\usepackage{soul}
\usepackage[font=footnotesize,labelfont=bf]{caption}
\usepackage[font=footnotesize,labelfont=bf]{subcaption}
\usepackage{authblk}
\usepackage{amsmath}
\usepackage{paralist}

\DeclareCaptionType{copyrightbox}

\makeatletter
\newcommand*{\inlineequation}[2][]{%
  \begingroup
    \refstepcounter{equation}%
    \ifx\\#1\\%
    \else
      \label{#1}%
    \fi
    \relpenalty=10000 %
    \binoppenalty=10000 %
    \ensuremath{%
      #2%
    }%
    ~\@eqnnum
  \endgroup
}
\makeatother

\algrenewcommand\alglinenumber[1]{\tiny #1:}

\algrenewcommand\algorithmicrequire{\textbf{Input}}
\algrenewcommand\algorithmicensure{\textbf{Output}}

\newtheorem{theorem}{Theorem}[section]
\newtheorem{lemma}[theorem]{Lemma}

\begin{document}

\title{Algorithms and Heuristics for Scalable Betweenness Centrality Computation on Multi-GPU Systems}

\author[1]{Flavio Vella \thanks{Part of this work performed while the author was at Scalable Parallel Computing Laboratory at ETH Z\"urich, Switzerland. E-mail: vella@di.uniroma1.it}}
\author[1]{Giancarlo Carbone}
\author[2]{Massimo Bernaschi}
\affil[1]{Department of Computer Science, Sapienza University of Rome, Italy}
\affil[2]{Istituto per le Applicazioni del Calcolo, IAC-CNR, Rome, Italy}
\date{}
\maketitle
\begin{abstract}
Betweenness Centrality (BC) is steadily growing in popularity as a metrics of the influence of a vertex in a graph.
The BC score of a vertex is proportional to the number of all-pairs-shortest-paths passing through it.
However, complete and exact BC computation for a large-scale graph is an extraordinary challenge that requires high performance computing techniques to provide results in a reasonable
amount of time.
Our approach combines bi-dimensional (2-D) decomposition of the graph and multi-level parallelism together with a suitable data-thread mapping that overcomes most of the difficulties caused by the irregularity of the computation on GPUs.
Furthermore, we propose novel heuristics which exploit the topology information of the graph in order to reduce time and space requirements of BC computation.
Experimental results on synthetic and real-world graphs show that the proposed techniques allow the BC computation of graphs which are too
large to fit in the memory of a single computational node along with a significant reduction of the computing time.

\end{abstract}


\section{Introduction}
\label{sec:introduction}

Graph analysis represents a fundamental tool in domains like the study of social networks \cite{wasserman1994social} and computational biology \cite{chen2009biomolecular}.
One of the main goals of graph analysis is to rank the nodes in a network according to a centrality measure.
In general, centrality measures play an important role in several graph applications including transport networks \cite{wang2011exploring} beyond the aforementioned social and biological networks \cite{freeman1977set, bullmore2009complex}.
One of the most popular metrics is the Betweenness Centrality (BC) \cite{freeman1977set}.
The fastest algorithm for calculating BC scores has $\mathcal{O}(nm)$ time-complexity and $\mathcal{O}(n+m)$ space-complexity (where $n$ is the number of vertices and $m$ is the number of edges) for unweighted graphs \cite{brandes2001faster}.
Therefore, the exact computation is infeasible for very large networks.
Several authors \cite{madduri2009faster, mclaughlin2014scalable, suriyuce2015} proposed to speedup the exact computation of BC scores by resorting to parallel processing of Brandes' algorithm.
However, in those solutions, the size of the graph is limited by the space complexity of Brandes' algorithm.
Pioneering approaches \cite{bulucc2011combinatorial, bernaschibc2015} overcome the memory limitations by distributing the graph among more computational nodes.
In particular, Bernaschi {\em et al.}, proposed the first fully distributed BC on clusters of Graphics Processing Units (GPU).
Recently, parallel architectures like Graphics Processing Units (GPU) have been successfully used in accelerating many irregular and low-arithmetic intensity applications like graph traversal-based algorithms, in which the control flow and memory access patterns are data-dependent \cite{burtscher2012quantitative, harish2007accelerating}.
Within this context, workload imbalance and uncoalesced memory accesses are major bottlenecks for GPUs.

Breadth-First-Search (BFS) represents a building-block for the solution of more sophisticated problems on unweighted graphs like minimum-cut \cite{cormen1990introduction}, ST-connectiviy \cite{Bernaschi2015145} and betweenness centrality as well \cite{brandes2001faster}.
Therefore parallel and distributed BC implementations can exploit innovative techniques for BFS.
In particular, GPU-based implementations require efficient and balanced threads-data mapping \cite{Merrill2012, Mastrostefano2013, jia2011edge}.
On distributed systems, there are several fast BFS implementations, e.g. \cite{beamer2013distributed} which combines a bottom-up and a top-down approach.
As for Multi-GPUs systems, in \cite{bisson2015} the authors proposed an efficient implementation of BFS for the Nvidia Kepler architecture.
Inter-node communication is considered a major bottleneck of early BFS implementations on distributed systems \cite{buluc2013graph, yoo2005scalable, lumsdaine2007challenges}.
Although advanced techniques overcome many of the difficulties (e.g., the use of a bit-mask improves the scalability \cite{Satish2012}), a distributed
BC implementation requires further/specific techniques due to Brandes' algorithm time-complexity and for the different memory requirements with respect to a simple BFS.
For instance, a distributed BC implementation requires information exchange about the shortest path so there is no benefit from using a bit-mask during the communication \cite{bernaschibc2015}.

The contributions of the present paper are manifold:
we present three complementary solutions for the computation of Betweenness Centrality for unweighted graphs based on three different level of parallelism.
We first describe: $i)$ an efficient algorithm for betweenness centrality computation that outperforms in most cases previous single GPU implementations by exploiting a threads-data mapping technique based on prefix-sum operations; $ii)$ a technique which mitigates the cost of prefix-sum;
we combine a communication-optimized fully distributed solution based on a two-dimensional (2-D) decomposition of the sparse adjacency matrix of the graph
together with concurrent Multi-Source search operations on
  Multi-GPU systems.
With the goal of reducing time and space requirements of BC
computation, two different heuristics are also presented.
The first one reduces the size of the graph in terms of both vertices and edges as well as the number of steps required to compute the exact BC score.
With respect to existing solutions, we extend the heuristics to graphs
with more connected components, providing also a fast distributed pre-processing algorithm.
The second heuristics, based on a novel approach, allows augmenting
the BC score of a 2-degree vertex from its adjacencies without performing Brandes'algorithm explicitly.
A theoretical insight about this result is also provided.

The rest of the paper is organized as follows:
in Section \ref{sec:backrelated}, we briefly describe Brandes' algorithm and recent results on exact betweenness computation with special focus on parallel and distributed implementations.
Our main contributions are presented in Section \ref{sec:gpubc}.
In particular, in Section \ref{sec:2degree} we provide the theoretical
foundations and the algorithm of the 2-degree heuristics, including some practical aspects.
In Section \ref{sec:exp}, we report comprehensive experimental results to validate our study.
Finally, in Section \ref{sec:conclusion} conclusions and future research directions are outlined.

\section{Background and related work}
\label{sec:backrelated}
Let $G = (V, E)$ be a graph representing a network composed by entities (vertices) and relations (edges), respectively.
Formally, let $G = (V, E)$  be a undirected and unweighted
graph with $ n = |V|$ vertices and $m = |E|$ unordered pairs $(u,v)$ such that $u,v \in V$ and $u \neq v$.
Since the graph is undirected, we consider $(u,v)$ and $(v,u)$ to be the same edge. The degree of a vertex  $deg(v)$ is the number of edges incident on it.
The shortest path between two vertices $s,t$ is a minimum-length sequence of unique vertices.
On unweighted graphs, a BFS solves the Single-Source-Shortest-Path problem in $\mathcal{O}(m)$ \cite{cormen1990introduction}.
The first formal definition of the betweenness centrality metrics was proposed in \cite{freeman1977set} (see also \cite{wasserman1994social} for further details).
Let $\sigma_{st}$ be the number of shortest paths between vertices $s,t$ whereas $\sigma_{st}(v)$ represents the number of those shortest paths that pass through $v$ with $s,t,v \in V$.
We define the pair-dependency on $v$ of a pair $s,t$, the ratio $\delta_{st}(v) = \frac{\sigma_{st}(v)}{\sigma_{st}}$.
The betweenness centrality of a vertex $v$ is defined as the sum of the pair-dependencies of all pairs on $v$,
\inlineequation[eq:bc]{
BC(v) = \displaystyle\sum_{s \neq t \neq v}^{} \delta_{st}(v)
}.
Before Brandes' work, a simple algorithm computed the BC score by solving the all-pairs-shortest-path problem and then by counting the paths.
That solution requires $\mathcal{O}(n^3)$ time by using the Floyd–Warshall algorithm and $\Theta(n^2)$ space for pair-dependencies.
In order to remove the explicit summation of all pair-dependencies and thus exploiting the natural sparsity of real-world graphs,
Brandes introduced the dependency of a vertex $v$ with respect to a source vertex $s$:
\begin{equation}
\label{eq:bc1}
 \delta_s(v) = \displaystyle\sum_{w:v \in pred(w)} \frac{\sigma_{sv}}{\sigma_{sw}}\cdot(1 + \delta_s(w))
\end{equation}
Formula \ref{eq:bc} can be re-defined as sum of dependencies:
\begin{equation}
\label{eq:bc2}
 BC(v) = \displaystyle\sum _{s \neq v} \delta_s(v)
\end{equation}

\begin{algorithm}[h]\scriptsize
\caption{Brandes' algorithm}
\begin{algorithmic}[1]
 \Require $G(V,E)$ \Comment G is unweighted graph
 \Ensure $BC[v], v \in V$
 \State $BC[v] \gets 0$
 \For{$s \in V$}
        \State $S \gets$ \texttt{empty stack}
        \State $Pred[v] \gets$ \texttt{NULL} $ \forall v \in V$
        \State $\sigma[v] \gets 0, \forall v \in V, \sigma[s] = 1$
        \State $d[v] \gets -1, \forall v \in V, d[s] = 0$
        \State $Q \gets$ \texttt{empty queue}
        \State \texttt{enqueue} $s \to Q$
        \While{Q not empty} \Comment Path counting via BFS
            \State \texttt{dequeue} $v \gets Q$
            \State \texttt{push} $v \to S$
            \ForAll{neighbor $w$ of $v$}
                \If{$d[w] < 0$} 
                \State \texttt{enqueue} $w \to Q$
                \State $d[w] \gets d[v] + 1$
                \EndIf
                \If{$d[w] = d[v] + 1$} 
                \State $\sigma[w] \gets \sigma[w] + \sigma[v]$
                \State \texttt{append} $v \to P[w]$
                \EndIf
            \EndFor
        \EndWhile
        \State $\delta[v] \gets 0, \forall v \in V$ \Comment Dependency
        \While{S not empty}
            \State \texttt{pop} $v \gets S$
             \For{$v \in Pred[w]$}
                  \State $\delta[v] \gets \delta[v] + \frac{\sigma[v]}{\sigma[w]} \times (1 + \delta[w])$
             \EndFor
             \If{$w \neq s$} \Comment Update BC
             \State $BC[w] \gets BC[w] + \delta[w]$
             \EndIf
        \EndWhile
 \EndFor

\end{algorithmic}
\label{alg:brandes}
\end{algorithm}

As a consequence, the BC score can be computed by solving the Single-Source-Shortest-Paths (SSSP) problem for each vertex in the graph.
To summarize, Brandes' algorithm, shown in Algorithm \ref{alg:brandes}, computes BC scores in $\mathcal{O}(nm)$
on unweighted graphs \cite{brandes2001faster} and consists in:
\begin{enumerate}
 \item computing the single source-shortest-path $\sigma$ from a single root vertex $s$ (lines $9-22$);
 \item summing all dependencies $\delta$ from $s$ (lines 25-29) and update BC score (line 30);
 \item repeating steps 1. and 2. for each vertex in $G$.
\end{enumerate}

\subsection{Related work}
Several authors have tackled the problem of speeding up the exact BC computation by parallelizing  Brandes' algorithm.
That approach requires a fast and memory-efficient traversal algorithm for unweighted graphs.
As mentioned above, BC computation on GPU suffers from both the irregular
access pattern and the workload unbalance due to traversal steps of the graph (counting of shortest paths and dependency accumulation).
Jia {\rm et al.} \cite{jia2011edge} evaluated two types of data-thread mapping: \textit{vertex-parallel} and \textit{edge-parallel}.
Briefly, the former approach assigns a thread to each vertex during graph traversal.
The number of edges traversed {\rm per} thread depends on the out-degree of the vertex assigned to each thread.
The difference in the out-degree among vertices causes a load imbalance among threads.
In particular, since the out-degree distribution of typical scale-free networks (like the social networks)
follows a power law \cite{Albert}, there is a severe load imbalance that explains the poor performance obtained with that approach on GPU systems.
The \textit{edge-parallel} approach solves that problem by assigning edges to threads during the frontier expansion.
However, this assignment of threads can also result in a waste of work because the edges that do not originate
from vertices in the current frontier do not need to be inspected.
The \textit{edge-parallel} approach is not well-suited for graphs with low average degree, as well as dense graphs \cite{jia2011edge}.
The vertex-based parallelism is affected by workload unbalance, whereas the edge-based parallelism uses
more memory and more atomic operations \cite{jia2011edge, Sariyuce2013}.
In \cite{Sariyuce2013, suriyuce2015} and \cite{mclaughlin2014scalable}, the authors proposed different
strategies in order to exploit the advantages of both methods.
In detail, Mclaughlin and Bader discussed two hybrid methods for the selection of the parallelization strategy.
Their sampling method performs on average $2.71$ times better than the
edge-parallel approach by Jia et al..
Sarıy\"{u}ce et al., in \cite{Sariyuce2013} and \cite{suriyuce2015}, introduced the vertex virtualization technique based on a relabeling of the data
structure (e.g., CSR, Compressed Sparse Row). Their solution is able to compute 32 concurrent BFS on a Nvidia Tesla K20 before decreasing in performance.
The technique replaces a high-degree vertex $\textit{v}$ with $n_v = \lceil \texttt{adj(v)} \rceil / \Delta $ virtual vertices having at most $\Delta$ neighbours.
In other words, the neighbours of high-degree vertices are divided (according to the input parameter $\Delta$) in several groups and each of them is assigned to a virtual vertex.
Vertex virtualization technique is not very effective for graphs with low average degree. Moreover, it requires a careful tuning of its parameters.
The authors also proposed a coarse-grained approach in which a single GPU executes multiple BFS at the same time with an increase of memory requirements.
Moreover in \cite{mclaughlin2015fast} an abstraction for processing multiple BFS on the GPU is provided. In that implementation, each source vertex is distributed across the Streaming Multiprocessors (SMs) of the GPU.
The threads within each warp process in parallel the edges outgoing from  the dequeued vertex collected by that warp.
Madduri et al. \cite{madduri2009faster} propose to check successors instead of predecessors in the dependency accumulation step.
In that way, the dependency accumulation procedure can start from one depth-level closer to the root vertex of the BFS tree and generally it
does not require atomic operations.
In \cite{green2013faster}, Green and Bader proposed a solution which reduces the memory requirements of local data structures from
$\mathcal{O}(m)$ to $\mathcal{O}(n)$ by discarding predecessors array on shared-memory system.

On distributed systems, betweenness computations can be parallelized in two ways: coarse- and fine-grained.
In the coarse-grained parallelism,  the entire graph and additional data structures are replicated so that each computing node has its own local copy.
Since each root vertex can be processed independently, each computing node processes a subset of the vertices of the graph.
At the end of the procedure, a \textit{Reduce} operation is also required to update the final BC scores.
For graphs that have a single connected component, the amount of work will be balanced among computational nodes.
In this case, a nearly perfect scaling can be expected \cite{mclaughlin2014scalable}.
However, this approach does not work in case of large scale graphs which cannot be stored in the memory of a single GPU.
On the contrary, in the fine-grained approach all processing units are
involved concurrently on the same computation starting from a single root vertex.
On distributed systems, this requires a partitioning of the graph and data structures among the computational nodes.
In \cite{edmonds2010space}, the authors proposed a space efficient distributed algorithm where the vertices are randomly assigned to each processor.
On unweighted synthetic graphs, the authors showed a satisfactory scalability up to 16 nodes.
Gunrock library also provides an implementation of Brandes' algorithm on a single Multi-GPU computing node \cite{gunrockmgpu}.
Their BC implementation is $2.5$ faster than the single GPU version proposed in \cite{gunrock2015} by exploiting 6 GPUs and 1-D partitioning.
In \cite{bulucc2011combinatorial} the authors adopt for the first time the 2-D partitioning for the betweenness computation.
Their solution solved the exact BC computation exploiting a Multi-Source BFS algorithm for the shortest path counting based on the linear algebra approach \cite{kepner2011graph}. In particular, the Multi-Source BFS is implemented as the multiplication of the transpose of the adjacency matrix of the graph ($M'$) with a rectangular matrix $F$, where each $i^{th}$ column of $F$ represents the current frontier of the $i^{th}$ concurrent BFSs.
However their solution did not exploit heuristics thus the performance are limited.
To the best of our knowledge, there are no solutions, based on
a linear algebra formulation, which exploit heuristics to speed-up the BC computation.
Bernaschi et al., in \cite{bernaschibc2015} proposed the first fully distributed BC on Multi-GPU systems. Their solution scales well up to 64 GPUs on Friendster graph \cite{slndc}.
The authors compared also two different partitioning strategies.
\subsection{Heuristics for Betweenness Centrality}
An exhaustive evaluation of betweenness centrality requires solving the SSSP problem starting from each vertex.
For large-scale graphs with millions of vertices, computing all SSSPs
is a formidable challenge.
Nevertheless, in some cases, the betweenness centrality of some sub-structures of the graph, or vertices  with specific properties, can be analytically computed with no need to execute Brandes' algorithm \cite{baglioni2012fast, sariyuce2013shattering, puzis2015topology}.
For example, Puzis et al. proposed two heuristics to speed-up the BC computation \cite{puzis2015topology}. The first one, contracts structurally-equivalent nodes (nodes that have the same neighbours) into one ``special'' node. The second heuristics relies on finding the biconnected components of the graph and contracting them as well.
The BADIOS framework, proposed in \cite{sariyuce2013shattering}, reduces the computation by shattering (bridges and articulation vertices) and compressing (side and identical vertices).
Moreover, focusing on compression based techniques, vertices with exactly one neighbor (1-degree vertices) have BC score 0, since they are endpoints and cannot be crossed by any shortest path.
As a matter of fact, a careful handling of 1-degree vertices improves overall performance of Brandes' algorithm:
a) by skipping the execution of Brandes' algorithm rooted from 1-degree vertices; b) by reducing the number of vertices to traverse.
Formally, let $G = (V, E)$  be an undirected and unweighted graph with $ n = |V|$ vertices and $m = |E|$ unordered pairs, let $(u,v) \in E : deg(u)$.
Since all the shortest paths terminating into a 1-degree vertex have to go through its neighbor,
the contribution $\delta_{sv}(w)$ could be not necessarily equal to 0.
From the algorithm point of view, 1-degree reduction extends Brandes' algorithm by adding a preprocessing procedure and by employing a different formulation for dependencies computation.
In detail, the preprocessing step computes  $\forall (u,v) \in E : deg(u) = 1$:
\begin{equation}
\label{eq:preprocessing}
\begin{split}
        \omega (v) & = \omega (v) + 1;\\  BC(v) & = BC(v)+ 2 \cdot (n - \omega(v) - 2)
\end{split}
\end{equation}
where $\omega (v)$ represents the contribution of $u$ to $v$ and initially is set equal to 0.
When a 1-degree vertex $u$ is detected, the value $\omega(v)$ of its neighbor $v$ is incremented, and $u$ is removed from the graph.
When $u$ is removed from the graph, the value $BC(v)$ needs to be updated in order to consider the contribution of paths starting
from all other vertices connected to $v$ and terminating in $u$. Notice that, $n$ does not correspond to the number of vertices in the graph, but to the number of vertices in the connected component of $v$.
After the preprocessing step, Brandes' algorithm is executed over the
residual graph $G'(V',E')$ obtained by the 1-degree removal procedure.
Concerning dependency accumulation, Formulas \ref{eq:bc1} and \ref{eq:bc2} can be re-defined as follows:
\begin{equation}
\label{eq:dependency1dg}
\begin{split}
  \delta_s(v) & = \sum_{w:v \in pred(w)} \frac{\sigma_{sv}}{\sigma_{sw}}(1 + \delta_s(w) + \omega(w)) \\
  BC(v) & = \sum _{s \neq v} \delta_s(w) \cdot (\omega(s) +1 )
\end{split}
\end{equation}
In \cite{Sariyuce2013,suriyuce2015} a GPU implementation of 1-degree reduction is provided.
However, the authors did not provide a general solution for graphs with more connected components.
As a matter of fact, their approach requires to compute the largest connected component of the input graph before the execution of the 1-degree reduction.
We propose a solution to overcome that limitation in Section \ref{sec:1degree}.
Moreover, in \cite{bernaschibc2015}, a distributed 1-degree reduction preprocessing based on 1-D partitioning is described and evaluated.
Their solution shows linear scalability up to 64 GPUs.
Another approach consists in re-using the shortest path tree from the vertex adjacent to a 1-degree vertex to cut down on computation \cite{bader2010large}.
Indeed the traversal step from a 1-degree vertex is not required if the shortest path tree from its adjacency is stored.
That solution does not require the preprocessing step, but, at the
same time, it does not take advantage of graph compression.

\subsection{2-D partitioning for traversal based algorithm}
\label{sec:partitioning}
A suitable decomposition of the graph is instrumental in order to achieve better performance and satisfactory scalability on distributed systems.
Different partitioning strategies can be adopted.
For example 1-D partitioning is a straightforward way of distributing the vertices of a graph \cite{Bernaschi:2015}.
It consists in assigning vertex $u_j,$ along with its outgoing edges, to the computing node $k$ according to the simple rule $k=j\%p$, where $p$ is the number of computing
nodes and $\%$ indicates the remainder of the integer division $j/p$.
However, for graph traversal based algorithms, 1-D partitioning suffers for poor scalability since it requires all-to-all communications among all the $p$ computing nodes \cite{yoo2005scalable, Buluc2011, Bernaschi2015}.
In \cite{yoo2005scalable, Buluc2011}, the authors proposed 2-D partitioning to reduce the communication cost.
2-D partitioning assumes that the processors are arranged as a bi-dimensional mesh having $R$ rows and $C$ columns.
The mesh is mapped onto the adjacency matrix $A_{N\times N}$ once horizontally and $C$ times vertically thus dividing the columns in $C$ blocks and the rows in $RC$ blocks.
Processor $p_{ij}$ handles all the edges in the blocks $(mR+i,~j)$, with $m=0,...,C-1$.
Vertices are divided into $RC$ blocks and processor $p_{ij}$ handles the block $jR+i$.
2-D partitioning can be summarized as follows:
\begin{inparaenum}
\item[{\em i)}] the edge lists of the vertices handled by each processor are partitioned among the processors in the same grid column;
\item[{\em ii)}] for each edge, the processor in charge of the destination vertex is in the same grid row\end{inparaenum}.
The traversal steps can be grouped in two major phases: $1)$ build the current frontier of vertices on each processor belonging to the same column of the mesh (\textit{expansion}); $2)$ exchange new discovered vertices involving the processor on the same row (\textit{folding}).
Let $p$ be the number of processors, the 1-D partitioning requires $\mathcal{O}(p)$ data transfers at each step, whereas the 2-D partitioning requires only $\mathcal{O}(\sqrt{p})$ communications since only the processors in one mesh-dimension are involved in the communication at the same time.

\section{Betweenness Centrality Computation on GPUs}
\label{sec:gpubc}

Our goal is to reduce the time-to-solution for the evaluation of BC when the size of the graph is such that even a parallel, shared-memory based, implementation is not a viable solution. We   believe that only the combination of a fast and scalable distributed solution with sophisticated heuristics enables the processing of large scale graphs.
Our Multi-GPU Betweenness Centrality (\textbf{MGBC}) algorithm consists in a sophisticated parallelization of Brandes' algorithm that exploits three different complementary levels of parallelism:
\begin{enumerate}
 \item at node-level: CUDA threads work on a subset of edges
   according to a suitable strategy of data-threads mapping.
 \item at cluster-level: a set of processors (or accelerators like
   GPUs) works concurrently following a graph partitioning strategy. At this level, the performance depends on the communication network as well.
 \item at subcluster-level: multiple sub-clusters work over replica
   of the same graph. Each sub-cluster performs the BC procedure on a subset of vertices concurrently to other sub-clusters. This further level of parallelism can be introduced since the betweenness equation is additive.
\end{enumerate}

The heuristics we present allow reducing the size of the graph in the traversal phase as well as skipping the BC procedure for vertices having specific features.
In particular, we introduce two complementary heuristics based on 1-degree and 2-degree vertices.

\subsection{Active-Edge Parallelism}
\label{sec:datamap}
In the present work, we extend the data-thread mapping approach originally introduced in \cite{bisson2015}.
That mapping strategy extends the edge-parallel approach by assigning a thread to each outgoing edge from the vertices in the current queue ($CQ$).
In this way, we do not need to inspect each edge in the graph like in the original edge-parallelism strategy.
To that purpose, it is necessary to count the total number of outgoing edges from the vertices in the frontier and
then map each vertex to its neighbors.
In detail, the degree of each vertex in the current frontier is stored into a contiguous array $CD$.
Then, a prefix-sum of the $CD$ array is performed.
At the end of the prefix-sum, $CD$ contains the information required to identify the predecessor vertex associated to the $i^{th}$ thread.
In order to identify the predecessor, a binary search over the $CD$ array is also required.
\begin{figure}[htbp!]
\begin{center}
\fbox{\includegraphics[scale=0.30]{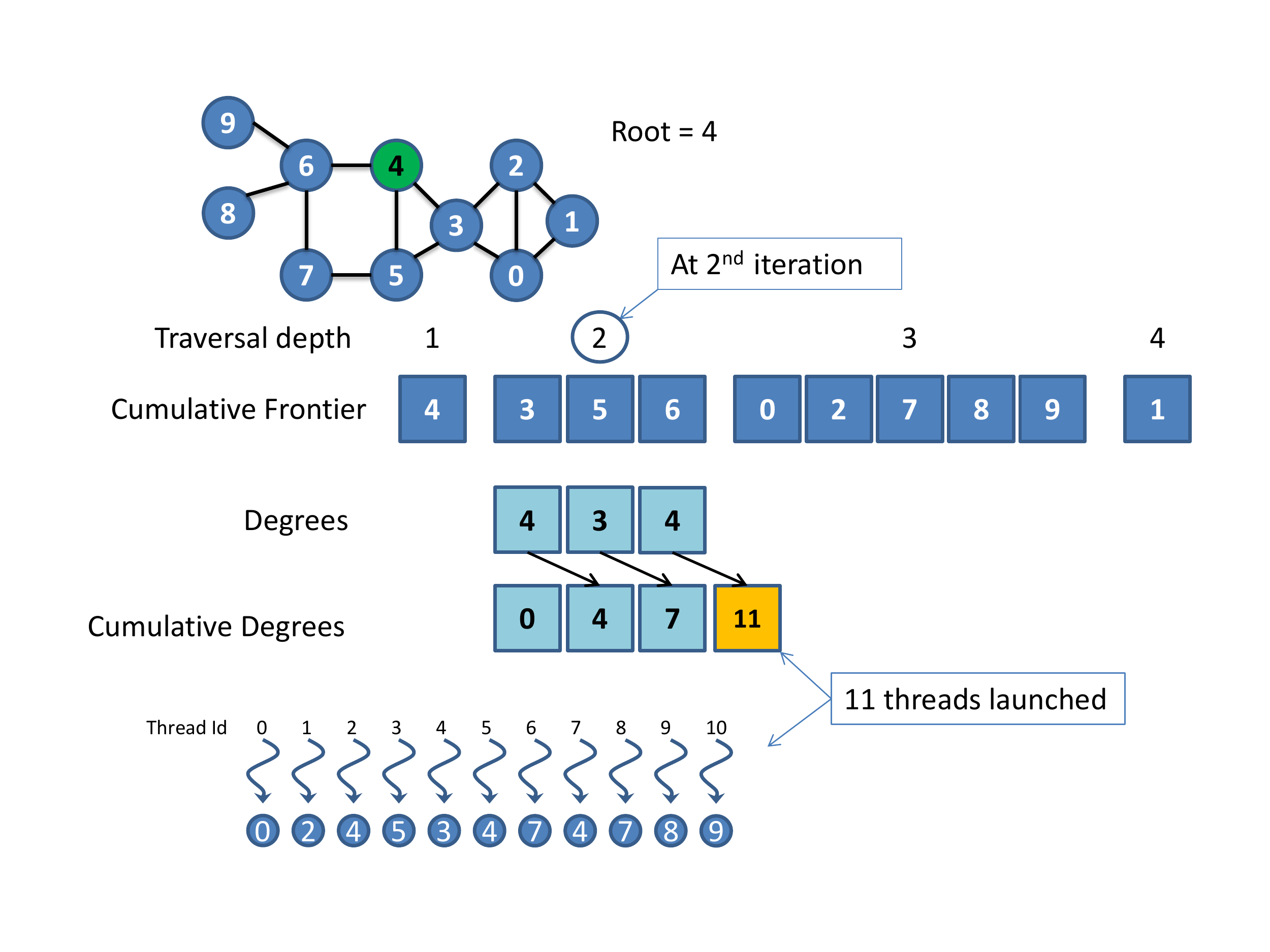}}
\end{center}
\caption{Example of data-thread mapping technique}
\label{fig:mapping}
\end{figure}
A simple example of that approach is illustrated in Figure \ref{fig:mapping}.
This mapping achieves a perfect load balancing among threads by introducing extra computation.
However, the prefix-sum and binary search operations may represent a significant overhead during the traversal steps.
We propose a strategy that reduces the cost of the prefix-sum and binary search.
In the Brandes' algorithm, given a root vertex, graph traversal occurs both in the shortest path counting and in the dependency accumulation procedure.
We observe that the latter operation is carried out along the BFS tree computed in the traversal step by visiting the same frontiers in reverse order.
Following a strategy based on active-edge parallelism, both steps will perform a scan operation on the degree of the same vertices in the same frontiers.
By storing and accumulating the offset array $CD$ during the shortest
path counting, we can avoid to perform again the prefix scan. This
solution allows reducing the computation time during dependency
accumulation by reading $CD$ stored in the previous step.
By exploiting the symmetry between forward and backward traversal step, the binary search results can be reused as well.
Obviously, this time-saving has an extra memory cost that is, at most, $\mathcal{O}(n)$.

\subsection{Betweenness Centrality on Multi-GPUs system}
Our Multi-GPU Betweenness Centrality algorithm (MGBC) is a novel parallelization of Brandes' algorithm.
Like Brandes' algorithm, MGBC is composed by three main steps:
$i)$ shortest paths counting, $ii)$ dependency accumulation and  $iii)$ update of BC scores.
Algorithm \ref{alg:spmulti} describes the shortest path counting procedure implemented in MGBC.
In lines 7-12, root vertex is enqueued and variables are initialized. 
At the beginning of each step, each processor has its own subset of
the frontier. According to the 2-D partitioning,
processors on the same column exchange frontier vertices (vertical communication), so all processors on the same column share the same frontier.
\begin{algorithm}[h]\scriptsize
\caption{Shortest Path Counting on Multi-GPU}
\begin{algorithmic}[1]
\Require $G(V,E)$ \Comment G is unweighted graph
\Require Processor $P_{ij}$
\Require Root vertex $s$
        \State $\sigma[v] \gets 0, \forall v \in V$
        \State $d[v] \gets -1, \forall v \in V$
        \State $Q \gets$ \texttt{empty queue}
                \State $lvl \gets 0$     \Comment{BFS level or depth}
                \State $nq \gets 1$
                \State $Q_{off}[0] \gets 0$

                \If{$s$ belongs to $P_{ij}$}
                        \State $\sigma[s] \gets 1$
                        \State $bmap[s] \gets 1$
                        \State $d[s] = 0$
                        \State \texttt{enqueue} $s \to Q$
                \EndIf

        \While{true}
                        \State $lvl \gets lvl + 1$
                        \State \texttt{gather} $Q$ and $\sigma$ from column $j$  \Comment{Vertical communication}
                \State $Q_{off}[lvl] \gets Q_{off}[lvl-1] + nq$
                        \State $nq \gets 0$

                        \State $Q_r \gets $ \textbf{expandFrt} $(lvl, bmap, Q, Q_{off}, d, \sigma)$  

                        \State \texttt{exchange} $Q_r$ and $\sigma$ for row $i$ \Comment{Horizontal communication}
                        \State \texttt{append} $Q_j \to Q$

                        \State \texttt{append} \textbf{updateFrt} $(lvl, bmap, Q, Qoff, d, \sigma) \to Q$      

                        \State $nq \gets $ \texttt{number of vertices added to } $Q$

                        \If{$nq = 0$ \textbf{for all} processors}
                                \State \textbf{break}
                        \EndIf

        \EndWhile

\end{algorithmic}
\label{alg:spmulti}
\end{algorithm}
During the frontier expansion (see Algorithm \ref{alg:expand}), new discovered vertices are marked as visited.
Their $\sigma$ values are updated by an atomic operation.
The edges belonging to other processors are communicated together with partial $\sigma$ values (horizontal communication).
At the end, the current frontier and $\sigma$ values are updated.
After shortest path counting, in our approach the depth array of each discovered vertex ($d$) is exchanged as well. This operation is performed once for each BC round between shortest-path counting and dependency accumulation phases.
\begin{algorithm}[h]\scriptsize
\caption{expandFrt}
\begin{algorithmic}[1]
 \Require $G(V,E)$ \Comment G is unweighted graph
 \Ensure $BC[v], v \in V$
        \ForAll{$v \in CQ$ in parallel} \Comment $CQ$ is the current frontier
                \ForAll{neighbor $w$ of $v$ in parallel}
                        \If{$bmap[w] = 0$}
                                \State $bmap[w] \gets 1$
                                \State $d[w] \gets lvl$
                                \State $r \gets$ row of $w$'s owner
                                \State \texttt{atomically enqueue} $w \to Q_r$
                        \EndIf

                        \If{$d[w] = lvl$}
                                \State \texttt{atomically} $\sigma[w] \gets \sigma[w] + \sigma[v]$
                        \EndIf

                \EndFor
        \EndFor

\end{algorithmic}
\label{alg:expand}
\end{algorithm}
In contrast to the 2-D BFS case, in the 2-D BC algorithm during each \textit{fold} phase the sigma values must be exchanged
reducing the scalability of the algorithm.
Moreover, in a straightforward implementation of Brandes' algorithm, the list of the predecessors of each vertex should be exchanged as well.
To avoid that, we discard the predecessors with the following benefits: $1)$ a reduction of the memory requirements of the local data structures from $\mathcal{O}(m)$ to $\mathcal{O}(n)$; $2)$ a reduction of \texttt{read} and \texttt{write} operations on GPUs memory. 
In distributed systems, due to $1)$, the communication cost decreases.
As a consequence, the modified distance BFS can be employed for the shortest path counting. 
By keeping track of local frontier expansion and combining that
information together with  the distance array, it is possible re-build
the predecessors/successors list with no additional communications among processors.
A similar technique has been adopted in \cite{green2013faster} on shared memory systems.
\newline Dependency accumulation is described in Algorithm \ref{alg:damulti}. Our approach is based on the checking successor technique \cite{madduri2009faster}.
Since leaves of the BFS tree do not have successors, the algorithm starts one level closer to the root.
As mentioned before, in line 1 of Algorithm \ref{alg:damulti}, both vertices' depth $d$ and  $\sigma$ are exchanged
among computing nodes in the same row.
\begin{algorithm}[h]\scriptsize
\caption{Dependency Accumulation on Multi-GPU}
\begin{algorithmic}[1]
\Require $G(V,E)$ \Comment G is unweighted graph
\Require Processor $P_{ij}$

                \State \texttt{exchange} $d$ and $\sigma$ for row $i$ \Comment{Horizontal communication}

        \State $\delta[v] \gets 0, \forall v \in V$

                \State $depth \gets lvl - 1$

        \While{depth $>$ 0}

                        \State \textbf{accumulateDep} $(depth, Q, Q_{off}, d, \sigma, \delta)$  \Comment{Accumulate dependencies}

                        \State \texttt{all reduce} $\delta$  among column $j$  \Comment{Vertical communication}

                        \State \textbf{updateDep} $(lvl, Q, Qoff, d, \sigma)$  \Comment{Update dependencies}

                        \State \texttt{exchange} $\delta$ among row $i$  \Comment{Horizontal communication}

                        \State $depth \gets depth - 1$

        \EndWhile

\end{algorithmic}
\label{alg:damulti}
\end{algorithm}
According to Brandes' algorithm, the dependency $\delta[w]$ is calculated by the shortest path count $\sigma[v]$ and
dependency value $\delta[v]$ of all its successors.
Each processor accumulates the local contributions to $\delta[w]$ for those successors for which it holds the edge $(w,v)$ (\textbf{accumulateDep} procedure on line 5).
All the local dependency contributions are then exchanged and summed by a reduce operation among the processors having the same index column of the mesh.
The final dependency value $\delta[w]$ is obtained multiplying the accumulated dependencies over $\sigma[w]$ (procedure \textbf{updateDep}).
Finally, $\delta[w]$ values are exchanged among processors on the same row (line 8) since they are required for the next iteration.
The procedure \textbf{accumulateDep} is described in Algorithm \ref{alg:accum}.
In detail, the algorithm first selects the vertices in the accumulated frontier $Q$ (line 1) to verify if their neighbors are successors (line 4).
Then another atomic operation is performed to update the local dependency $\delta[w]$.
\begin{algorithm}[h]\scriptsize
\caption{AccumulateDep}
\begin{algorithmic}[1]

        \State $CQ \gets Q[Q_{off}[depth]] ... Q[Q_{off}[depth-1]]$

        \ForAll{$w \in CQ$ in parallel}
                \ForAll{neighbor $v$ of $w$ in parallel}
                        \If {$d[v] = d[w] + 1$}
                                \State \texttt{atomically} $\delta[w] \gets \frac{1+\delta[v]}{\sigma[v]}$
                        \EndIf
                \EndFor
        \EndFor

\end{algorithmic}
\label{alg:accum}
\end{algorithm}

\begin{figure}[]
\begin{center}
\includegraphics[width=0.65\textwidth]{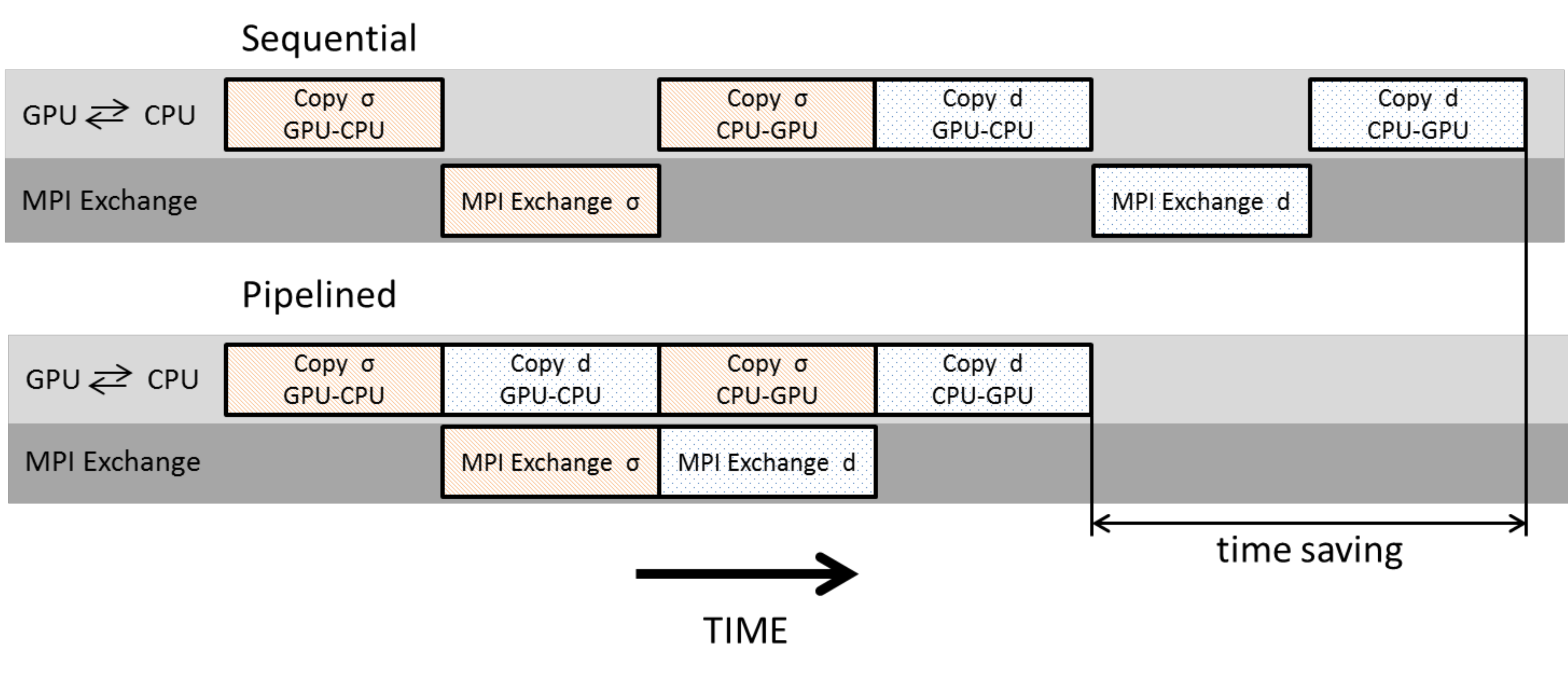}
\end{center}
\caption{Overlapping of GPU - CPU data transfer with MPI communication.}
\label{fig:overlap}
\end{figure}

Finally, the proposed distributed algorithm allows for the overlap of MPI communication and CPU-GPU data transfer. 
Although Nvidia provides several techniques to reduce communication
overhead such as GPUDirect RDMA \cite{gpurdma}, we adopt a simple
overlap mechanism
between two consecutive communications, whereby the cost of the communication through the PCI bus can be hidden.
In particular, right after the shortest path counting phase, both the distance vector $d$ and $\sigma$ values are exchanged among processors in the same grid row.
Since the computation is totally delegated to GPU, usually two consecutive independent communications comply with the following pattern:
\begin{enumerate}
 \item synchronous-copy of $\sigma$ from GPU to CPU;
 \item exchange of $\sigma$ among processors in the same grid row;
 \item synchronous-copy of $\sigma$ from CPU to GPU.
 \item synchronous-copy of $d$ from GPU to CPU;
 \item exchange of $d$ among processors in the same grid row;
 \item synchronous-copy of $d$ from CPU to GPU.
\end{enumerate}
In this naive pattern, data transfer procedure ends after six synchronous steps.
However, by exploiting Cuda Asynchronous Copy operations and Cuda Streams, the two communications can be completed in four steps (see Figure \ref{fig:overlap}):
\begin{enumerate}
 \item asynchronous-copy of $\sigma$ from GPU to CPU; asynchronous-copy of $d$ from GPU to CPU;
 \item exchange $\sigma$ among processors in the same grid row;
 \item asynchronous-copy of $\sigma$ from CPU to GPU; exchange $d$ among processors in the same grid row;
 \item asynchronous-copy of $d$ from CPU to GPU.
\end{enumerate}

\subsection{Sub-clustering}

\begin{figure}[htbp!]
\begin{center}
\includegraphics[scale=0.30]{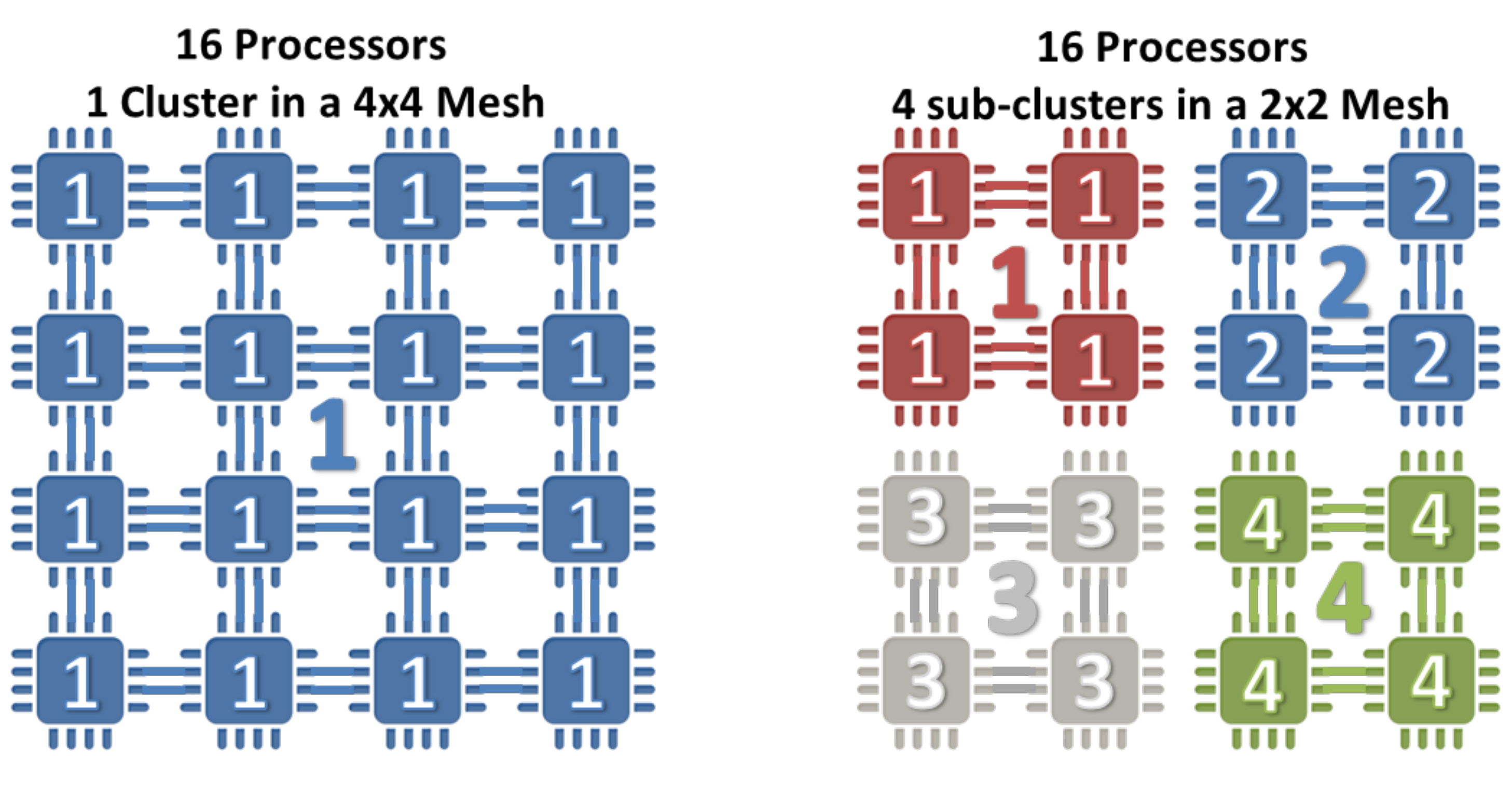}
\end{center}
\caption{Sub-clustering. On the left side the configuration ($p=16$,
  $fd=1$ and $fr=1$) enables a pure fine-grained strategy. On the
  right side, a sub-cluster configuration with $p=16$, $fd=4$ and $fr=4$.}
\label{fig:cluster}
\end{figure}

A Multi-Source approach for the BC computation offers a significant speed-up on a single-GPU, provided that extra-memory (for example for the replication of $\sigma$ and $\delta$ arrays) is available, as reported in \cite{suriyuce2015}.
In addition, on distributed systems the replication of data-structures may increase the communication among computing nodes and increase the synchronization points.
For example, the approach adopted in \cite{bulucc2011combinatorial}
encapsulates three levels of parallelism: columns of F provide
  parallelism over starting vertices, columns of $M'$ and rows of F provide
  parallelism over the vertices in each frontier. Finally, rows of
  $M'$ encapsulates edge (adjacency) parallelism of each frontier
  vertex. However all the processors in the mesh are involved in the communication during traversal steps.
Therefore, to the best of our knowledge, using the single-GPU Multi-Source approach as a basis for a fully distributed BC algorithm does not appear the best option.
On the other hand, on distributed systems, a coarse-grained approach
enables to obtain a great speed-up by replicating the data structures among computing nodes in order to work on multiple vertices at the same time.
As above mentioned, this approach limits the maximum size of the graph that can be processed (i.e., Twitter graph \cite{Kwak10www} cannot be
stored in the memory of a single GPU).
However, it is possible to obtain a significant improvement of performance by combining fine- and coarse-grained approaches at cluster level abstraction.
Within this context, we propose a solution to combine graph distribution and graph replication on a Multi-GPUs system.
Although the present work is focused on BC and Multi-GPUs systems, the approach is more general and can be followed for most problems (e.g., diameter computation, all-pairs-shortest-paths, transitive closure, etc...) that require multiple, independent breadth-first searches on graphs too large to fit in a single computing node.
A set of processors is split into sub-clusters.
Each sub-cluster, in turn, is organized as a bi-dimensional grid of processors.
Processing nodes in the same sub-cluster work at the fine-grained level:
the graph is distributed among the nodes according to a 2-D partitioning, and partial BC values are calculated starting from a subset of vertices.
Independent sub-clusters work at the coarse-grained level: the whole graph and additional data structures are replicated in each sub-cluster.
In the end, a reduce operation updates the final BC scores.
Even if the amount of work in each sub-cluster can be different when processing graphs with multiple connected
components, with the sub-clusters solution it is possible to take advantage of both fine- and coarse-grained approach (see Section \ref{sec:expmultisub}).
Let $p$ be the number of processors available/requested in the
cluster, and let $fd$ be the factor of graph distribution (indicating the size of the mesh of the sub-cluster).
The factor of replication of the graph ($fr$) is defined by $fr = \frac{p}{fd}$ and, in our implementation, it determines the number of sub-clusters.
A simple example is shown in Figure \ref{fig:cluster}.
On the contrary to existing solutions, like \cite{bulucc2011combinatorial}, which involves all $p$ processors in the communication, sub-clustering
technique involves only $fd$ processors in a subcluster during
traversal steps (expect for the final reduction operation).
Furthermore our approach is not limited to a 2-D partitioning so other partitioning strategies can be adopted.
Both the $fd$ and $fr$ factors must be taken into account to achieve best performance.
Concerning practical aspects, we implement this solution by creating a hierarchy among processes managed by different MPI communicators.

\subsection{Heuristics}
\label{sec:heuristic}
\subsubsection{1-Degree Reduction}
\label{sec:1degree}
In this Section, we discuss our algorithm for the removal of 1-degree vertices\footnote{For the sake of simplicity we do not remove tree vertices from the graph by calling repeatedly the preprocessing (tree vertices removal).}$^{,}$\footnote{The preprocessing is implemented only on CPU.}.
Unlike previous approaches, we provide a distributed preprocessing
algorithm described by the pseudo-code in Algorithm \ref{alg:1dgpreprocessing}.
One-degree reduction requires to identify vertices with degree one and this task is easier to accomplish if each vertex,
along with all its edges, is stored on the same processor.
This can be easily obtained with a 1-D partitioning (see Section \ref{sec:partitioning}).
First, the edges are sorted by the antecedent vertex $u$ and processed sequentially:
when a 1-degree vertex $u$ is discovered, $\omega[v]$ is incremented
and the edge $(u,v)$ is added to the list $R$ of the removed edges.
Otherwise, all the edges from $u$ are appended to the new edge list $E'$ of the residual graph.
In an undirected graph, for each edge $(u,v)$ the symmetric edge
$(v,u)$ must be removed as well.
The contribution of 1-degree vertices to BC scores cannot be computed during preprocessing since we support graphs with multiple connected components, on the contrary to previous solutions.
By observing the formula  $BC(v) = BC(v) + 2 \cdot (n - \omega(v) -2)$, we already highlighted that $n$ corresponds to the number of vertices in the same connected component of $v$, including 1-degree vertices. 
For any vertex $s$, we can compute $n_s$, the number of vertices of its connected component, during shortest paths counting.
When a new vertex $v$ is discovered during graph traversal from root vertex $s$, $n_s$ is updated as follows: $n_s = n_s + \omega[v]$.
Computing $n_s$ is required whenever $\omega[s]\neq 0$, in other words, only if vertex $s$ is connected to a 1-degree vertex.
There are two alternatives for the computation of $n_s$: $i)$ using atomic operations during
shortest paths counting; $ii)$ using a parallel reduction of the
distances array before the update of the betweenness centrality score.
In both cases, the procedure should not consider the contribution $\omega(v)$ of unvisited vertices.
As to the performance, the best solution depends on the cost of atomic operations.
Finally, since our approach does not require information about the connected components of the graph. the computing time of the preprocessing step decreases.

\begin{algorithm}[h]\scriptsize
\caption{1-Degree Preprocessing}
\begin{algorithmic}[1]
        \Ensure $\omega[v]$, $G'(V',E')$
        \State $R \gets$ \texttt{empty List}
        \State $E'  \gets$ \texttt{empty List}
        \If{$u \bmod \#P = P_i: (u,v) \in E$} \Comment $P_i$ is processor $i^{th}$
            \State \texttt{assign (u,v) to} $E_i$
        \EndIf
        \State \texttt{sorting $E_i$ by} $u$
        \For{$(u,v) \in E_i$}
                \If {$\not\exists (w,z) \in E_i: u = w$} \Comment (w,z) the successor or predecessor in E
                        \State  \texttt{append} $(v,u) \to R$
                        \State $\omega[v] = \omega[v] + 1$
                \Else
                \State  \texttt{append} $(u,v)\to E'$
                \EndIf
        \EndFor

\end{algorithmic}
\label{alg:1dgpreprocessing}
\end{algorithm}

\subsubsection{Augmenting BC of degree-bounded vertex}
\label{sec:2degree}
In this Section, we propose a new technique based on dynamic
programming, to compute the BC score of 2-degree vertices without executing Brandes'algorithm from them explicitly.
In contrast to the 1-degree reduction or other techniques which modify the topology of the graph, the 2-degree heuristics exploits the information of the shortest-path tree of the two neighbors to derive both shortest-path tree and dependency of the 2-degree vertex.
A similar intuition is barely sketched in \cite{madduri2008high}. The authors proved that it is possible to build the shortest path tree from an arbitrary vertex in the graph without re-traversing the graph when the shortest-path trees from all its adjacencies are known. However, they did not provide neither an algorithm nor related results.

The following notation is used in the rest of the Section.
We denote with $lvl_s(v)$  the discovery depth or level or unary distance of $v$ during a traversal step from a source vertex $s \neq v$.
Let $c$ be a vertex with $deg(c)=2$ and $a$ and $b$ its own neighbors. We also use the symbols $frt_s$ and $frt_s^{k}$ to denote the BFS tree (or set of frontiers) of a vertex $s$ and the set of vertices discovered at level $k$ respectively.

Our goal is to compute the shortest path and the dependencies of a
2-degree vertex by re-using shrewdly the information provided by the
execution of Brandes' algorithm for its adjacencies. To that purpose,
we need to determine:
\begin{enumerate}
 \item the $frt_c$ from the frontiers of its own adjacencies;
 \item the number of the shortest paths of each vertex from $c$;
 \item the dependencies of $c$.
\end{enumerate}

Concerning the first point, the key idea behind the 2-degree heuristics is that the frontiers of a 2-degree vertex can be built by merging the frontiers of the two neighbors.
To do that, the BFS trees of its own adjacencies must be stored.
Furthermore, it is apparent that the shortest paths from $c$ to a vertex $v \in frt_c$ must pass through either $a$, $b$ or both; indeed we may
determine the relation between the level at which a vertex is
discovered starting from $c$ and the level of the same vertex discovered starting from the 2-degree neighbors.

\begin{lemma}
 \label{theo:shortest-sigma}
Let $c$ be a 2-degree vertex, and let $a$ and $b$ be neighbors of $c$
such that $a$, $b$ and $c$ $\in V$ in an unweighted graph $G=(V,E)$.
Each $v$ vertex in the frontiers of $a$ and $b$ obeys to the following rules:
$i)$ $v$ is $\in frt_c$.
$ii)$ $v$ is discovered in $frt_c$ at level $lvl_c(v) = min \{ lvl_a(v),lvl_b(v)\} + 1$
\end{lemma}

The number of shortest paths passing through vertex $v$ in $frt_c$ depends on which shortest-path is followed when $v$ is discovered at $lvl_c(v)$, a path through $a$ or $b$.
In other words, $\sigma_c(v)$ is equal to $\sigma_a(v)$ iff $lvl_a(v) < lvl_b(v)$; likewise, $\sigma_c(v)$ is equal to $\sigma_b(v)$ iff $lvl_b(v) < lvl_a(v)$.
If $v$ is discovered at the same level from both $a$ and $b$ (i.e., $lvl_a(v) = lvl_b(v)$), then $\sigma_c(v)$ is defined by the shortest paths
passing via $a$ and $b$.

More in detail, we first recall the Bellman's observation.
\begin{lemma}{(Bellman criterion)}
 \label{theo:bellman}
A vertex $v \in V$ lies on a shortest path between vertices $s$, $t \in V$, if and only if $d(s, t) = d(s, v) + d(v, t)$.
\end{lemma}

By properly applying Bellman criterion we find that:
\begin{equation}
\label{eq:2-degree-shortest}
  \sigma_c(v) =
  \begin{cases}
    \sigma_a(v)  & \quad \text{if } lvl_a(v) < lvl_b(v)\\
    \sigma_b(v)  & \quad \text{if } lvl_a(v) > lvl_b(v)\\
    \sigma_a(v)+\sigma_b(v) & \quad  \text{if } lvl_a(v) = lvl_b(v)
  \end{cases}
\end{equation}
We prove Lemma \ref{theo:shortest-sigma} by induction.
At level 1, $frt_c^{1}$ is composed by $a$ and $b$ by definition of 2-degree vertex.
At level 2, $frt_c^2$ is composed by $frt_a^1 \bigcup frt_b^1$.
At level 1, $frt_a^{1}$ is composed by $c$ (by definition of $a$), a set of vertices $v \neq c \neq b$ iff $\exists (a,v) \in E$ (case 1) and  $b$ if exists an edge $(a,b)$ (case 2).
Likewise, $frt_b^{1}$ is defined by $c$, a set of vertices $v \neq c
\neq a$ iff $\exists (b,v) \in E$ (case 1) and $a$ if it exists an edge $(a,b)$ (case 2).
Let omit the case 2 ($a,b \notin E$).
If a generic vertex $v$ is discovered at $lvl^{i-th}$ from $a$, then
the path to reach $v$ from $b$ is longer or equal at most.
With respect to $c$, $v$ is reachable from the path passing through $a$ or $b$.
A naive implementation is described by the pseudo-code in Algorithm
\ref{alg:2-degree-shortest}. The algorithm makes it possible to get rid of the computation of the shortest path from $c$ only, so the dependency accumulation is still required.

\begin{algorithm}[h]\scriptsize
\caption{Shortest-path tree computation of a 2-degree vertex from its own neighbors}
\begin{algorithmic}[1]
        \Require $G(V,E)$, $sigma_a[]$, $sigma_b[]$, $lvl_a[]$ and $lvl_b[]$
        \Ensure $\sigma_c[]$, $lvl_c[]$
        \State $sigma_c[v]\gets 0, lvl_c[v] \gets \infty \quad  \forall v \in V$
        \ForAll{$v \in V$ in parallel}
                        \If {$lvl_a[v] = lvl_b[v]$}
                                \State  $\sigma_c[v] \gets \sigma_a[v]+ \sigma_b[v]$
                                \State $lvl_c[v] \gets lvl_a[v] + 1$
                        \EndIf
                        \If {$lvl_a[v] < lvl_b[v]$}
                                \State  $\sigma_c[v] \gets \sigma_a[v]$
                                \State $lvl_c[v] \gets lvl_a[v] + 1$
                        \Else
                                \State  $\sigma_c[v] \gets \sigma_b[v]$
                                \State $lvl_c[v] \gets lvl_b[v] + 1$
                        \EndIf
        \EndFor

\end{algorithmic}
\label{alg:2-degree-shortest}
\end{algorithm}

The BFS tree rooted in $c$ can be derived by simply sorting $lvl_c$. 
This solution only saves the time for the graph traversal.
We may achieve a greater benefit if betweenness contributions from $c$ are directly added on-the-fly while the dependency accumulation steps for its two neighbors $a$ and $b$ are performed.
This solution avoids both the execution of Algorithm \ref{alg:brandes} from $c$ and the explicit evaluation of $lvl_c$ and $\sigma_c$.
As explained before, the BC contributions $\delta_s$ of a vertex $s$ are computed recursively by re-traversing the BFS tree rooted in $s$ according to Formula \ref{eq:bc1}.
As a matter of fact, the $\delta_s$ at each level depends on the contributions at the deeper level.
The first problem to be considered is when the vertices contributions of $c$ should be added to $\delta_c$ since the order of visit may be different between its own neighbours.
This is accomplished by modifying the Brandes procedure so that dependency accumulation steps for $a$, $b$ and $c$ are performed together ''level by level``. During this step, the frontiers of $a$ and $b$ are dynamically merged (without storing them in a new BFS tree of $c$ explicitly) and contributions of $c$ dependencies are added as well.
We call this technique ``Dynamic Merging of Frontiers (DMF)".
In detail, Algorithm \ref{alg:2-degree-brandes} and Algorithm \ref{alg:2-degree-accumulation} modify the procedure described in Algorithm \ref{alg:brandes} at lines (24 - 28) by implementing DMF.
We first compute $\sigma_a$, $lvl_a$, $\sigma_b$ and $lvl_b$ (i.e., by performing the procedure described in Algorithm \ref{alg:expand}).
At line 1, the deeper BFS tree between $a$ and $b$ is evaluated.
The vertices in the leaves of $a$ and $b$ contribute to the $\delta_c$ iff their discovered level is the same for both.
For instance, let $w$ be a vertex belonging to the leaves of the BFS tree of $a$.
It may be discovered two levels before by $b$ (if $(a,b) \notin V$ ). In this case, the contribution of the predecessors of $w$ should be taken into account in $\delta_c(w)$ when $w$ is visited in the dependency accumulation of $b$. Moreover, we have to consider the shortest path tree of $b$ in the dependency accumulation formula.
When both current depths of the BFS trees are synchronized, the
procedure simultaneously computes, level-by-level, the dependencies for $a$, $b$ and $c$.
Algorithm \ref{alg:2-degree-accumulation} shows the dependency accumulation of a child of 2-degree vertex $c$ according to Formula \ref{eq:2-degree-shortest}.
In detail, within each iteration of the dependency accumulation, for
each vertex in the frontier of $a$, we calculate the dependency
accumulation as in the original algorithm but we check, in addition,
if the vertex should be considered for $c$. We do the same for each vertex in the frontier of $b$.
Notice that, when a predecessor $v$ of $w$ is discovered at the same
level in $a$ and $b$, the $\sigma_c(v)$ is defined for both $\sigma_a(v)$ and $\sigma_b(v)$ (line 6).
Like in Algorithm \ref{alg:accum}, the procedure exploits atomic operations to update $\delta_c$.
Finally, we can conclude with the following result.
\begin{theorem}
\label{theo:shortest-lvl}
Let $c$ be a 2-degree vertex, and let $a$ and $b$ be neighbors of $c$ such that $a$,$b$ and $c$ $\in V$ in an unweighted graph $G=(V,E)$.
The shortest path tree of $c$ can be derived iff the levels of each vertex discovered in BFS trees rooted $a$ and $b$ is given respectively.
\end{theorem}

\begin{algorithm}[h]\scriptsize
\caption{Dependency Accumulation steps based on Dynamic Merging of Frontiers}
\begin{algorithmic}[1]
        \Require $G(V,E)$, $sigma_a[]$, $sigma_b[]$, $lvl_a[]$ and $lvl_b[]$
        \Ensure $\delta_a[]$, $\delta_b[]$ and $\delta_c[]$
        \State $depth \gets $ \texttt{max} $\{depth_a, depth_b \}$
        \While {$depth > 0$}
        \If {$depth = depth_a$}
        \State \texttt{DependencyAccumulation-2degree} $(DepInfo_a, \sigma_a, lvl_a, \sigma_b, lvl_b)$ \Comment $DepInfo_a$ denotes the information required in Alg. \ref{alg:damulti} at line 5 related a vertex $a$.
        \State $depth_a$\texttt{--}
        \EndIf
        \If {$depth = depth_b$}
        \State \texttt{DependencyAccumulation-2degree} $(Q_a, \sigma_a, lvl_a, \sigma_b, lvl_b)$
        \State $depth_b$\texttt{--}
        \EndIf
        \State $depth$\texttt{--}
        \EndWhile
\end{algorithmic}
\label{alg:2-degree-brandes}
\end{algorithm}

\begin{algorithm}[h]\scriptsize
\caption{Augmenting the betweenness centrality accumulation from a left-child of a degree-2 vertex}
\begin{algorithmic}[1]
        \Require $DepInfo_a, Q_a, \sigma_a, lvl_a, \sigma_b, lvl_b, \delta_c$
        \State $CQ_a \gets Q_a[Q_{off}[depth]] ... Q_a[Q_{off}[depth_a-1]]$
        \ForAll{$w \in CQ_a$ in parallel}
                \ForAll{neighbor $v$ of $w$ in parallel}
                        \If {$d[v] = d[w] + 1$}
                                \State \texttt{atomically} $\delta_a[w] \gets \frac{1+\delta_a[v]}{\sigma_a[v]}$
                                \If {$lvl_a[v] = lvl_b[v]$} {atomically} $\delta_c[w] \gets \frac{1+\delta_c[v]}{\sigma_a[v]+\sigma_b[v]}$
                                \EndIf
                                \If {$lvl_a[v] < lvl_b[v]$} {atomically} $\delta_c[w] \gets \frac{1+\delta_c[v]}{\sigma_a[v]}$
                                \EndIf
                        \EndIf
                \EndFor
        \EndFor

\end{algorithmic}
\label{alg:2-degree-accumulation}
\end{algorithm}

The effectiveness of the 2-degree heuristics depends on the order in which the vertices are processed in the main loop of Brandes' algorithm.
When a 2-degree vertex is selected for the execution, first we have to perform the shortest paths counting steps from its own adjacencies.
At the same time, a 2-degree vertex should be processed together with its two neighbors.
Moreover, we cannot execute Brandes' procedure of a generic vertex $v$ without knowing if $v$ is a neighbor of a 2-degree vertex.
The solution proposed allows computing the dependency values of $a,b$ and $2-degree$ vertex $c$ in a single computation by concurrent execution of the dependency accumulation of $a$ and $b$ level-by-level.
This happens when the adjacencies of a 2-degree do not belong to the adjacencies of other 2-degree vertices.
As a matter of fact, we cannot solve all 2-degree vertices by applying Algorithm \ref{alg:2-degree-brandes} even if the graph is composed by 2-degree vertices only.
For instance, let $C=(V,E)$ be a cycle graph where $|V|= n$ and each vertex has degree 2. The algorithm computes the BC score of, at most, $\frac{n}{2}$ or $\lfloor \frac{n}{2} \rfloor-1$ (if $n$ is odd) vertices without performing Brandes algorithm explicitly.
As to memory requirements, the heuristics requires $\mathcal{O}(n)$ extra memory-space since both $\sigma_a$ and $lvl_a$ depends on the number of vertices of the graph.
In the present work, we do not address the problem to find out the minimal set of vertices for which we need to store the shortest path trees.
However, for experimental validation, we simply check if a vertex $v$ is a child of a 2-degree. If this occurs, the algorithm performs shortest paths counting from both $v$ and the other child of its own predecessor.
On the other hand, if $v$ is a 2-degree vertex, we execute the
shortest paths counting of its own adjacencies and then Algorithm
\ref{alg:2-degree-brandes} is used in order to derive the contribution
of $v$ to the BC.

\section{Experimental Results}
\label{sec:exp}
We first compare MGBC with other implementations on a single GPU.
Actually, most of them do not offer full support for a distributed
Multi-GPU configuration. Some of them working on distributed systems,
like \cite{mclaughlin2014scalable}, support only coarse-grained
parallelism, where each GPU works independently on a replica of the same graph.
All those solutions cannot be used for very large graphs, like Friendster or Twitter \cite{Kwak10www} since those graphs do not fit in the memory of a single system.
On distributed systems, weak and strong scalability experiments are performed in order to evaluate the ratio between computation and communication on different kind of graphs.
We also show the impact of the optimizations techniques here proposed on the performance.
Then, we measure the speedup provided by heuristics with respect  to our base (heuristics-free) implementation.
In particular, we evaluate the speedup of the 2-degree heuristics and its impact on graphs having a long diameter like road networks.

\subsection{Evaluation Platforms and Data Sets}
Numerical experiments have been carried out on two different systems: {\em Piz Daint} at Centro Svizzero di Calcolo Scientifico (CSCS) and {\em Drake}, a server equipped with four K80s GPU available at National Research Council of Italy.
Daint is a hybrid Cray XC30 system with 5272 computing nodes interconnected by an Aries network with Dragonfly topology. Each node is powered by an Intel Xeon E5-2670 CPU and a NVIDIA Tesla K20X GPU and is equipped with 32 GB of DDR3 host memory and 6 GB of DDR5 GPU memory.
The code has been generated using the GNU C compiler version 4.8.2, CUDA C compiler version 6.5 and Cray MPICH version 7.2.2 on Piz Daint and OpenMPI 1.8.4 on Drake.
We employ the exclusive scan implemented in the Thrust Library \cite{thrust}. The code uses 32-bit data structures except for graph generation.
We usually report the time (in seconds) for total BC computation.
Sometimes, in order to compare our results with those reported in state-of-the-art literature, we also present the {\em traversed edges per second} (TEPS) value as defined in the following formulas:
\begin{equation}
 TEPS_{bc} = \frac{m \times n}{t}
\end{equation}
where $n$ is the number of vertices or a subset of them, $m$ is the number of (undirected) edges, and $t$ is the execution time of the BC computation\footnote{We do not consider disconnected vertices.}.
However, for very large graphs we measure the time only for a representative subset of source vertices\footnote{Source vertices are selected randomly among not isolated vertices.}.
In this case, the expected time for the whole graph is derived.
We measured the performance for both R-MAT \cite{chakrabarti2004r} and real-world graphs (see Table \ref{tab:dataset}) \cite{slndc}.
As for R-MAT graphs, the number of vertices is defined by a {\tt scale} factor and it is equal to $2^{\texttt{scale}}$.
The edge factor parameter (EF) defines the number of edges as follows: $2^{\texttt{scale}}\times \texttt{EF}$.
We generate R-MAT graphs using parameters a, b, c, and d equal to 0.57, 0.19, 0.19, 0.05 respectively.
\newcommand{\specialcell}[2][c]{\begin{tabular}[#1]{@{}c@{}}#2\end{tabular}}

\begin{table*}[t]\scriptsize
\centering
\begin{tabular}{|c|c|c|c|c|c|c|c|}
\hline
\textbf{Graph} & \textbf{SCALE} & \textbf{EF} &\textbf{1-degree}  &\textbf{d}\\

\hline\hline

com-amazon      & $18.35$       & $2.76$   & $4.68$ &  $44$\\
com-youtube     & $20.11$       & $2.6$    & $53.00$ &  $20$ \\
RoadNet-CA      & $20.91$       & $1.41$   & $16.27$ &  $849$ \\
RoadNet-PA      & $20.05$       & $1.41$   & $17.13$ &  $786$\\
com-LiveJournal & $21.93$       & $8.67$   & $19.2$ &  $17$\\
com-Orkut       & $21.55$       & $38.14$  & $2.21$ &  $9$\\
Friendster      & $25.97$       & $27.53$  & $1.2$ &  $32$\\
Twitter         & $25.3$       & $35.25$  & $4.5$ &  $18$\\
\hline
\end{tabular}
\caption{Features of the real-world graphs used for the tests
  (\textbf{d} represents the diameter)}
\label{tab:dataset}
\end{table*}

\begin{table*}\scriptsize
\centering
\begin{tabular}{|c|c|c|c|c|c|}
\hline
\textbf{Graph} & \textbf{\specialcell{Mclaugh- \\ lin}} & \textbf{\specialcell{Sarıy\"{u}ce \\ mode-2}} & \textbf{\specialcell{Sarıy\"{u}ce \\ mode-4}} & \textbf{Gunrock} & \textbf{MGBC}\\
\hline\hline
RoadNet-CA      &   $\boldsymbol{0.067}$          & $0.371$    & $0.184$      & $0.298$       & $0.085$\\
RoadNet-PA      &   $\boldsymbol{0.035}$          & $0.210$    & $0.114$      & $0.212$       & $0.071$\\
com-Amazon      &   $0.008$          & $0.009$    & $0.006$ ($0.007$)           & $nt$       & $\boldsymbol{0.005}$\\
com-LiveJournal &   $0.210$          & $0.143$    & $\boldsymbol{0.084}$      & $nt$         & $0.100$\\
com-Orkut       &   $0.552$          & $0.358$    & $\boldsymbol{0.256}$      & $nt$         & $0.314$\\
\hline
\end{tabular}
\caption{Comparison with other single GPU implementations on real-world graphs.}
\label{tab:res-single}
\end{table*}

\subsection{Single-GPU}
The Single-GPU implementation is obtained from MGBC by turning off
network and related host-device communications.
We compare our solution on single GPU (without any heuristics or optimization) to those proposed in Mclaughlin and Bader \cite{mclaughlin2014scalable}, Sarıy\"{u}ce et al. \cite{Sariyuce2013} and Gunrock \cite{gunrock2015} on Drake system. The implementation described in \cite{suriyuce2015} is not available for the comparison.
In Table \ref{tab:res-single}\footnote{The acronym \texttt{nt} stands for ``execution does not terminate''.}, we report the mean time (in seconds) for each implementation. Since other codes do not allow a random selection of the vertices, the mean time is computed over the first 10000 vertices of the biggest connected component.
Concerning Sarıy\"{u}ce's implementations, we evaluated two of their data-mapping strategies. The first one, called \textit{mode-2} employs edge-based GPU parallelism, whereas the second one, \textit{mode-4}, uses virtual-vertices with stride access \cite{Sariyuce2013}.
Concerning mode-4, we also report in parentheses the virtualization time.
Experiments show that the hybrid approach of McLaughlin performs better than others on graphs with a pretty small edge factor and long diameter, like road networks.
Instead on other kind of graphs, the performance of their approach is not satisfactory.
On the other hand, the vertex-virtualization technique achieves very good performance on more dense graphs. However, such approach requires an \textit{a priori} tuning of the virtual-vertex parameter. By changing the virtualization parameter the performance may decrease.
Although the design is focused on distributed systems, our BC implementation achieves pretty good performance without requiring any specific tuning.

\subsection{Multi-GPU and Sub-Clustering}
\label{sec:expmultisub}
\begin{figure*}
\minipage{0.48\textwidth}
\includegraphics[width=.95\linewidth]{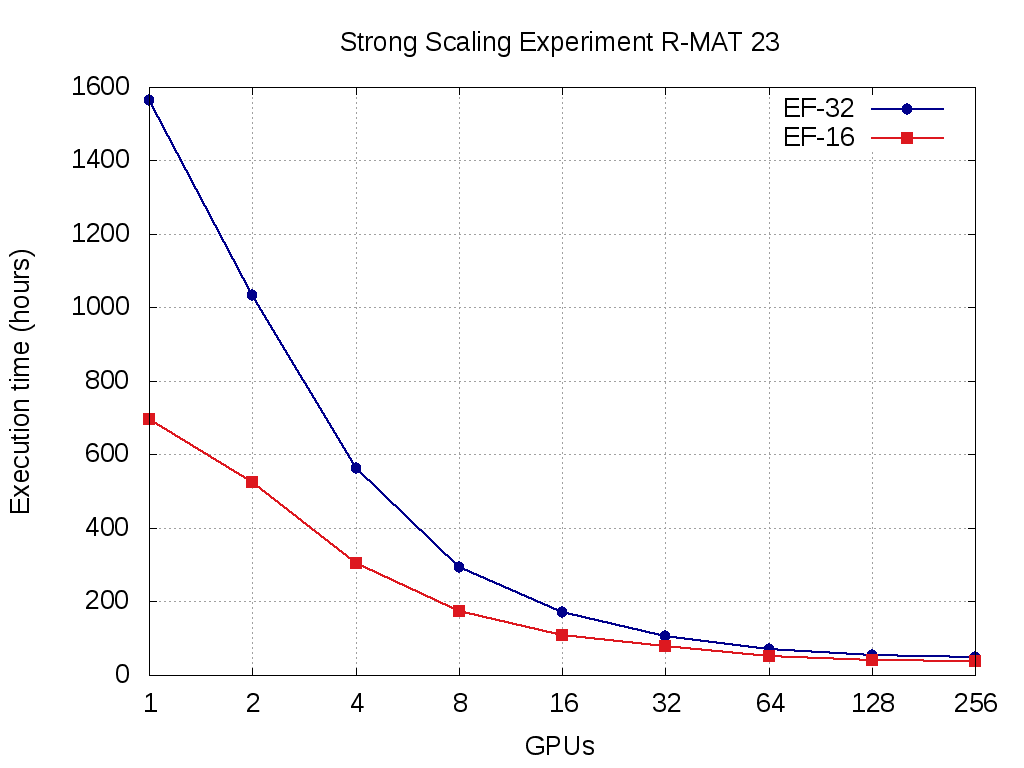}
        \caption{Strong scaling experiments for R-MAT graphs with SCALE 23 and EF 16 and 32.}
        \label{fig:strong1}
\endminipage\hfill
\minipage{0.48\textwidth}
\includegraphics[width=.95\linewidth]{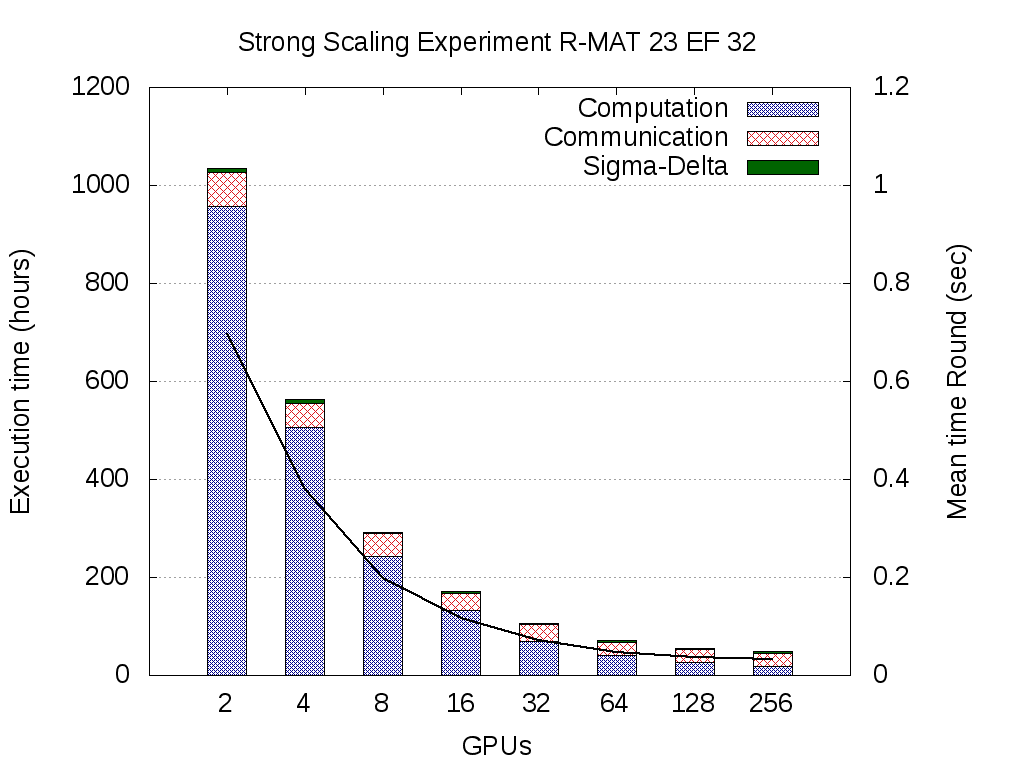}
        \caption{Strong scaling experiments for R-MAT graphs with SCALE 23 and EF 32.}
        \label{fig:strong2}
\endminipage\hfill

\end{figure*}

\begin{figure}[htbp!]
\centering
                  \includegraphics[scale=0.29]{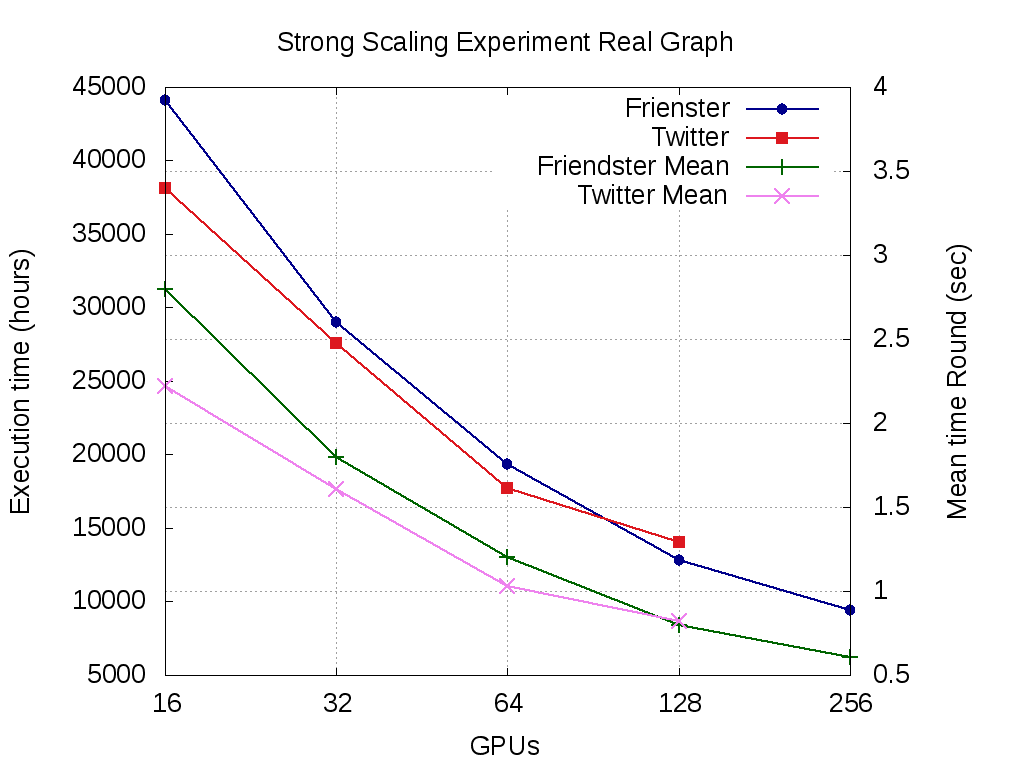}
                  \caption{Strong scaling experiments for Twitter and Friendster graphs.}
                  \label{fig:strongreal}
\end{figure}

\begin{figure*}[htbp!]
\minipage{0.48\textwidth}
        \includegraphics[width=1.0\linewidth]{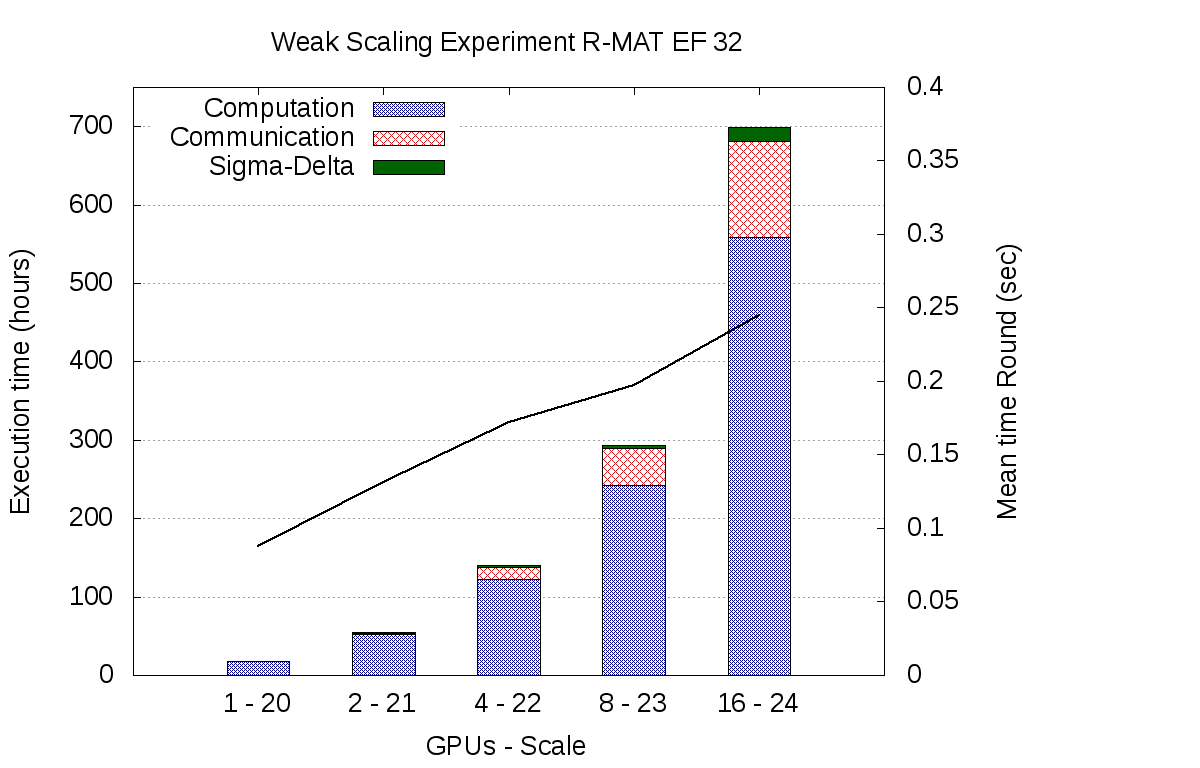}
        \caption{Weak scaling experiments for R-MAT graphs with EF 32 up to SCALE 24.}
        \label{fig:weak}
\endminipage\hfill
\minipage{0.48\textwidth}
        \includegraphics[width=1.0\linewidth]{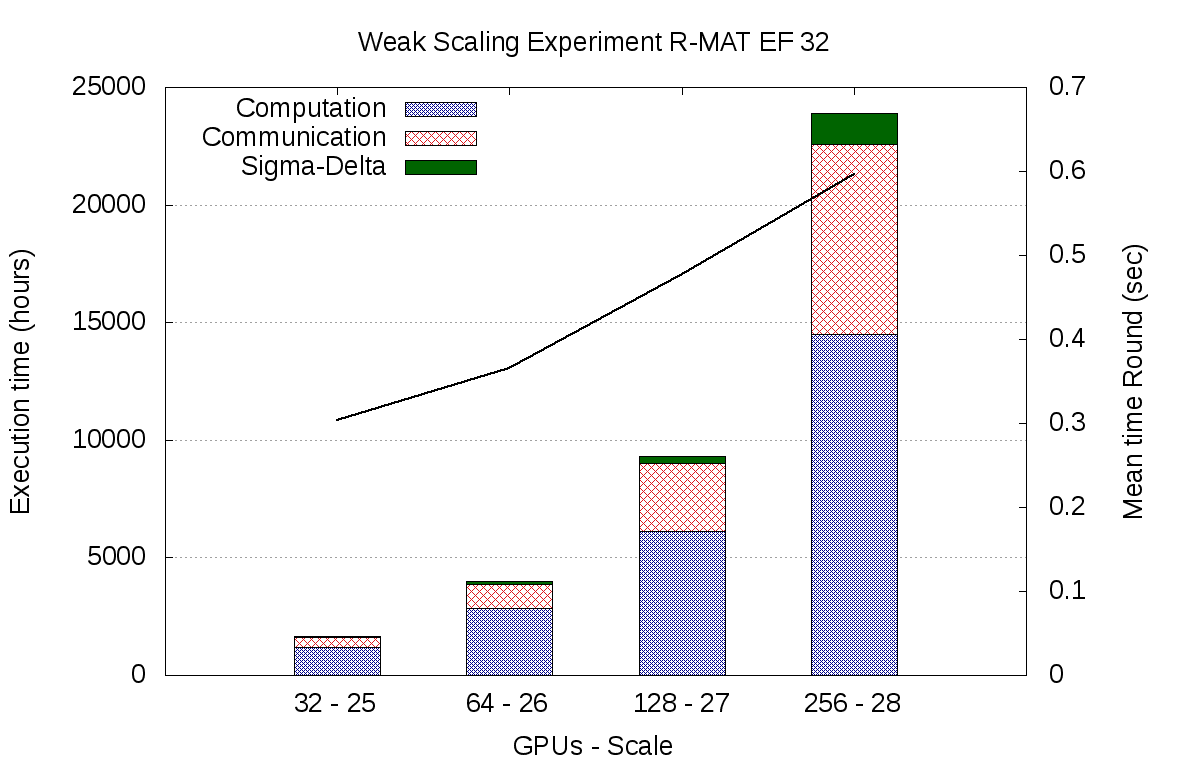}
        \caption{Weak scaling experiments for R-MAT graphs with EF 32 from SCALE 25 to 28.}
        \label{fig:weak2}
\endminipage \hfill
\end{figure*}

We evaluated performance on fixed-size graphs while increasing computational resources on Piz Daint (strong scaling experiments).
During these experiments, we performed 10000 BC computations without using neither heuristics nor the prefix-sum optimization.
We studied the scalability of MGBC on both R-MAT and real-world graphs.
In particular, Figure \ref{fig:strong1} shows the strong scalability for R-MAT graphs at SCALE 23 and two different edge factor, 16 and 32 respectively.
Our solution has a very good scaling up to 128 GPUs with EF 32.
Moving from 1 to 2 nodes, there is only a $\sim 40\%$ of improvement due to communication overhead.
In Figure \ref{fig:strong2}, we report a breakdown of the total time in computation, communication and $sigma$-$delta$ communication which measures the time spent in exchanging $\sigma$, $d$ and $\delta$ among the processors.
From 2 to 8 nodes, the scalability is almost linear. In those cases,
the communication represents a small fraction of the total time ($\sim
16\%$). From 16 to 32 nodes, we observe that the computation decreases
linearly whereas the communication remains almost the same. By employing 64 GPUs, both computation and communications decrease, however the computation represents one third of the total time. At $128$ nodes the computation and the communication are the same, then with more than $256$ GPUs, the communication dominates the computation and the algorithm does not scale anymore.
The $sigma$-$delta$ communications in the worst case (with $256$ GPUs) represent $\sim 9\%$ of the total time.
The black curve denotes the mean time for a BC round.
Concerning real graphs, the strong scaling for {\em Friendster} and {\em Twitter} graphs is also evaluated in Figure \ref{fig:strongreal}.
The mean time of a BC round is figured out by looking at the y2 axis.
Notice that the minimum number of GPUs required to store the graph is 16 for both graphs.
Although we observe a good scalability (up to $256$ GPUs for {\em Friendster}), the mean time of a BC round is still pretty high (i.e., $0.601$ seconds). As  a consequence, the exact computation of the betweenness centrality for both graphs is not feasible in a reasonable amount of time.
\newline Figure \ref{fig:weak}, \ref{fig:weak2} illustrate the performance of MGBC for graphs that a single GPU can not handle due to memory limits in weak scale experiments.
Although the amount of data is the same for each GPU, the time required to compute BC is not constant.
In particular, for R-MAT graphs with different EFs and SCALE, the time increases linearly from SCALE 20 up to SCALE 28 with EF 32.
The mean time for a BC round at SCALE 28 is $0.590$ seconds ($29$ GTEPS). To the best of our knowledge, there are not studies of BC on R-MAT graphs with SCALE greater than 24.

\begin{figure*}[t!]
    \centering
    \begin{subfigure}[t]{0.32\textwidth}
        \centering
        \includegraphics[scale=0.17]{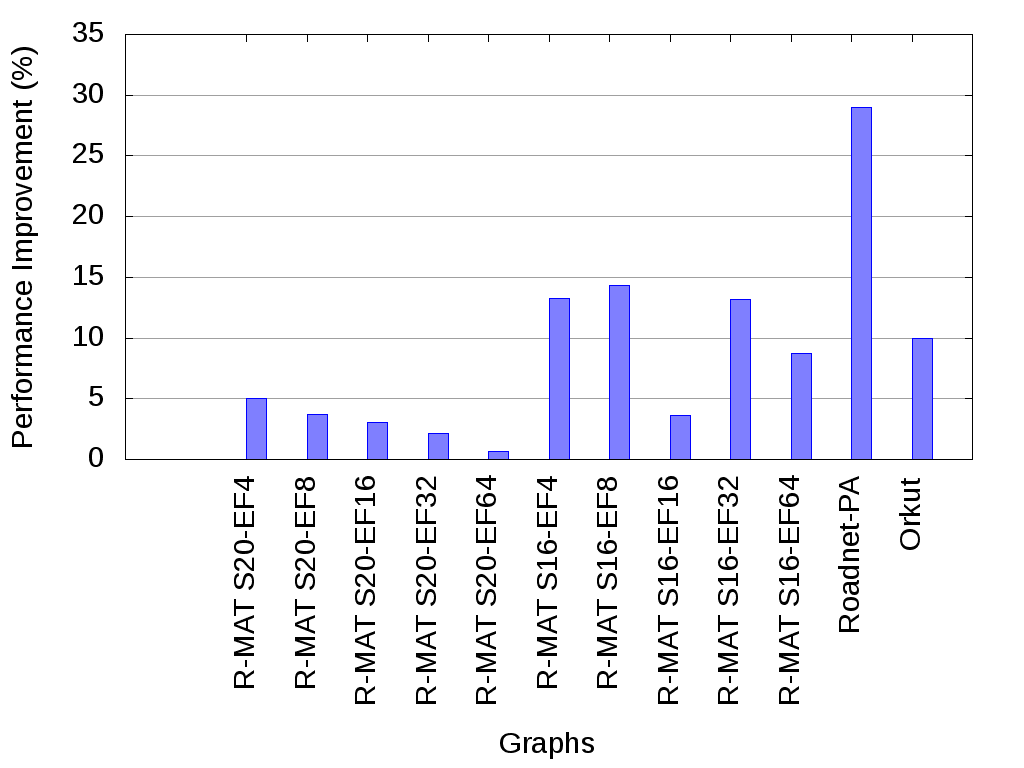}
        \label{fig:prefix}
        \caption{Impact of the prefix-sum.}
    \end{subfigure}
    \begin{subfigure}[t]{0.32\textwidth}
        \centering
        \includegraphics[scale=0.17]{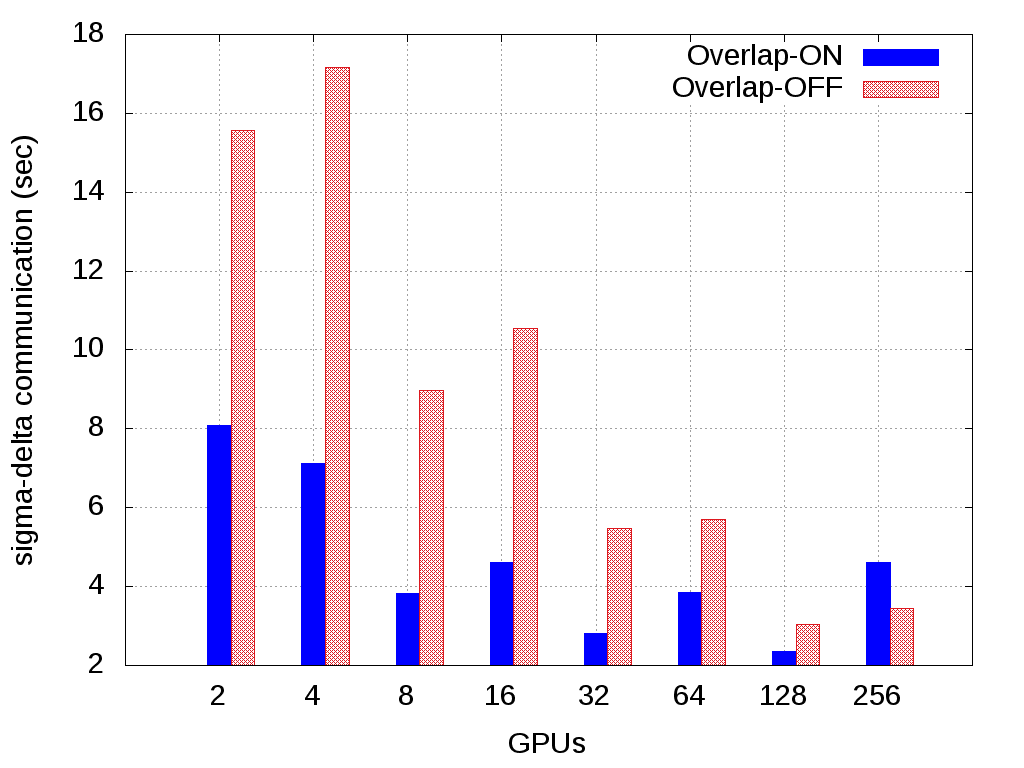}
        \caption{Impact of the overlap on R-MAT graph SCALE 23 EF 32.}
    \end{subfigure}
    \begin{subfigure}[t]{0.32\textwidth}
        \centering
        \includegraphics[scale=0.17]{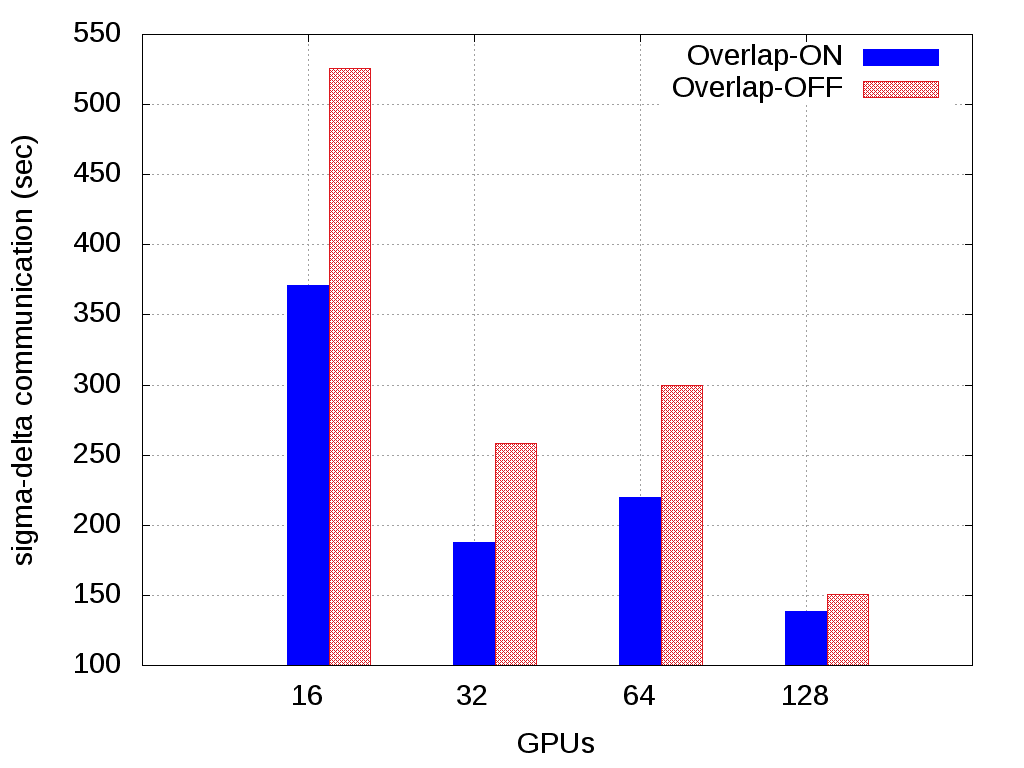}
        \caption{Impact of the overlap on Twitter graph.}
    \end{subfigure}
    \caption{Impact of optimizations on Single and Multi-GPU systems.}
    \label{fig:opt}
\end{figure*}

Multi-GPU implementation enables to handle very large graphs, however the overall time required can still be quite long.
For example, the full evaluation of BC for the Orkut graph on a single GPU requires $\sim 250$ hours \footnote{The time reported is obtained on a single node on Piz Daint.}.
However, by combining coarse- and fine-grained parallelism a substantial time reduction can be obtained.
Table \ref{tab:cluster} shows the total time required to compute BC for the Orkut graph when the number $fr$ increases. The $fd$ parameter used for the experiments is equal to $2$.
In this case the graph is distributed among a mesh of $2x1$ processors.
We also evaluated the performance with $fd=4$ (the processors are
organized a in 2x2 mesh).
Fixing $p = 256$ ($fr=64$), the time required for the full BC
computation of Orkut graph is $2.3$ hours. Therefore in this case, by
fixing $p$, a smaller factor of distribution offers the best performance.
Concerning the scalability, the sub-clustering technique requires that each sub-cluster had a balanced workload.
The BC scores of local copies are accumulated for all of the GPUs on each sub-cluster.
Finally, the scores at sub-cluster level are aggregated into the global BC scores by a reduce operation.
On {\em Orkut} graph, the workload among sub-clusters is balanced since {\em Orkut} unveils only one connected component.
\newline Finally, in Figure \ref{fig:opt}, we show the impact of the prefix-sum-free optimization and overlap technique.
To evaluate the impact of the prefix-sum optimization, we compared both implementations on graphs with different diameter and density.
Figure \ref{fig:opt} (a) shows the performance increment both on R-MAT and real graphs. In general, the improvement is more significant on graphs with long diameter since the prefix-sum is performed for each level and those graphs require many iterations.
The R-MAT graphs are characterized by short diameter, furthermore when the graph becomes denser the prefix-sum implemented in the Thrust library is more efficient since it tends to achieve the maximum throughput. More details on the performance of scan functions are reported in \cite{bruce2015}.
By observing the results on R-MAT with SCALE 16, we obtain a $14\%$
improvement due mainly to the low throughput of the prefix-sum implemented in Thrust.
On the other hand, RoadNet-PA graph is characterized by a long diameter and low density. In this (best) case, our technique offers the highest improvement ($\sim 30\%$).
The experiment on the {\em Orkut} graph (EF $\sim 38$ and diameter 9) represents the case where the prefix-sum is efficient but its cost is
relevant. As a matter of fact, the maximum cost of the prefix-sum is achieved when the algorithm traverses the levels (middle) where the maximum number of vertices is discovered.
In the latter case we obtain $10\%$ of improvement.

In order to evaluate the overlap technique, we compare the result of
strong scaling experiments previously reported with the result
obtained when the overlap is off.
In the strong scaling experiment, the amount of data stored in a single node varies.
In this way, we can evaluate the overall time for $sigma$-$delta$ exchange decreasing the cost of the host-device communication, by increasing the number of GPUs.
Figure \ref{fig:opt} (b) and (c) remark the effectiveness of our solution both on synthetic and real-world graphs.
In particular, the communication of $sigma$-$delta$ can be reduced by
a factor of $2.5$ when the overlap is enabled.
Notice that when the communication dominates the computation, the overlap benefit decreases.

\subsection{Heuristics}
\begin{figure*}[htbp!]
\minipage{0.48\textwidth}
\includegraphics[width=.92\linewidth]{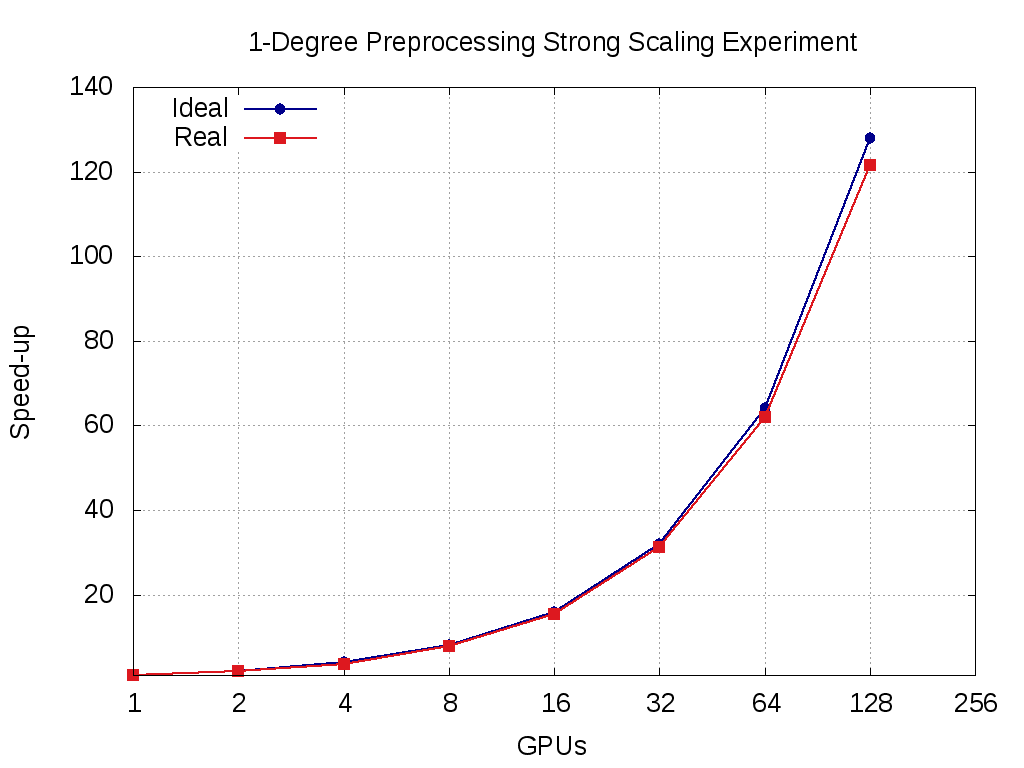}
        \caption{Strong scaling experiment of the preprocessing
          algorithm for a R-MAT graph with SCALE 23 and EF 32.}
        \label{fig:h1preproc}
\endminipage\hfill
\minipage{0.48\textwidth}
\includegraphics[width=.92\linewidth]{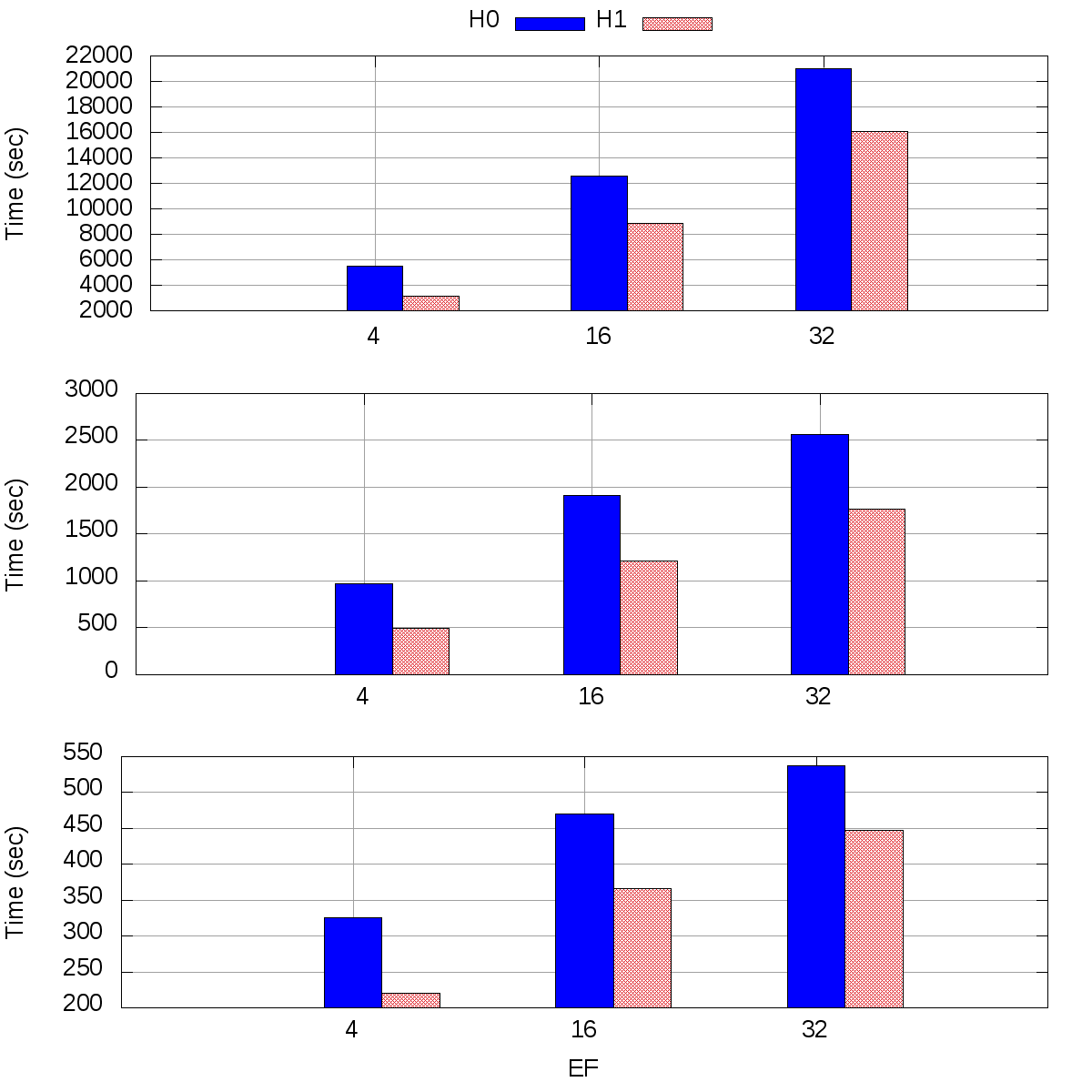}
        \caption{Impact of 1-degree reduction on a R-MAT graph with SCALE 20. The bars shows the time in seconds of the computation (top), communication (middle) and overlap (bottom) respectively.}
        \label{fig:h1}
\endminipage\hfill 
\end{figure*}

\begin{figure}
 \centering
    \includegraphics[scale=0.36]{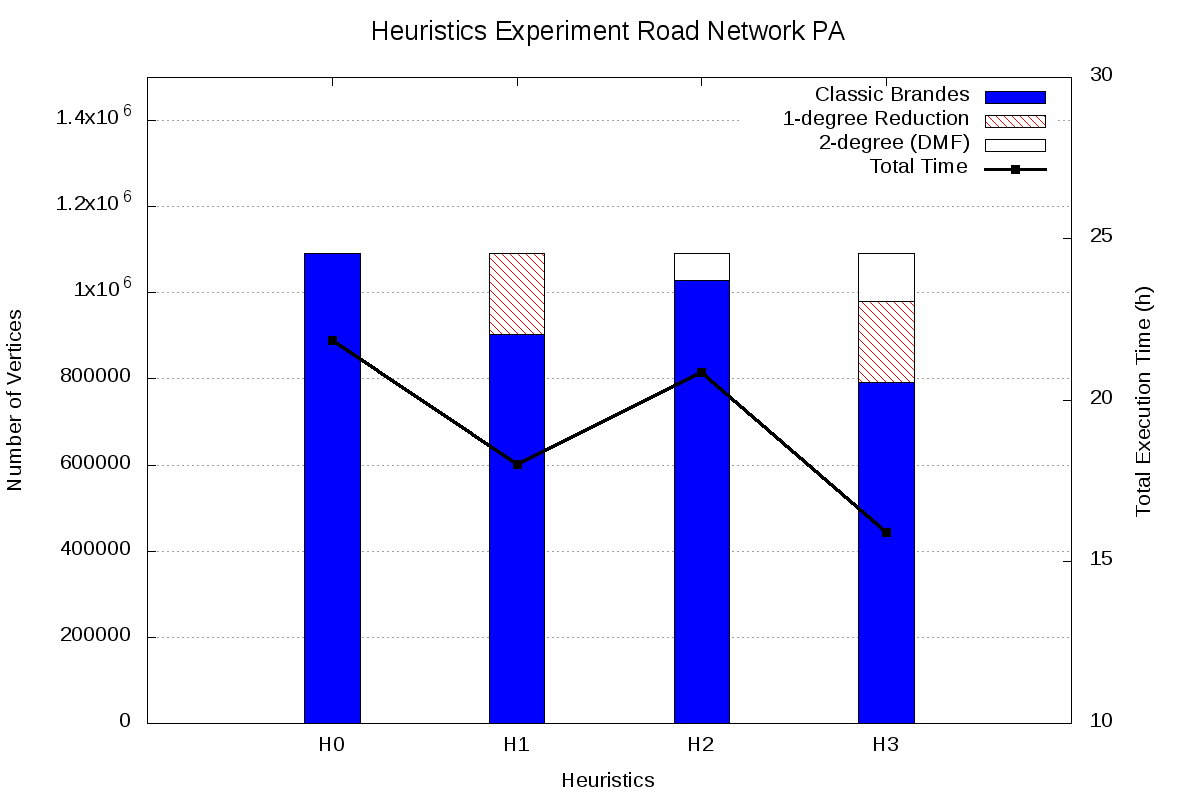}
    \caption{Heuristics comparison on RoadNet-PA.}
    \label{fig:h2}
\end{figure}

For the 1-degree reduction heuristics, we evaluated, first of all, the strong scalability of the preprocessing step.
Figure \ref{fig:h1preproc} illustrates the strong scaling of Algorithm \ref{alg:1dgpreprocessing} applied to a R-MAT graph with SCALE 22 and EF 16 on Piz Daint.
The algorithm exhibits a near-linear speedup suggesting that the communication does not represent a bottleneck during the preprocessing step.
The experiments reported below have been performed on the Drake system.
Concerning synthetic graphs, we computed the BC scores of all vertices of a R-MAT graph with SCALE 20 and different EFs exploiting a 2x2 grid of GPUs.
More in detail, Table \ref{tab:1degree} shows the mean time of an iteration of MGBC\footnote{The mean time is computed considering only connected vertices.}, the total time and the preprocessing time when the 1-degree heuristics is applied.
On a R-MAT graph with EF 16, the preprocessing takes less than $0.02 \%$ of the total time offering an increment of performance of $30\%$ compared to the execution with 1-degree off.
A more significant improvement can be achieved when the edge factor decreases since the number of 1-degree vertices increases. For
example, the execution of MGBC with 1-degree reduction on the {\em
  com-youtube} graph is $\sim 3$ times faster than an
execution with 1-degree reduction off.
On the contrary to previous works which show only the speed-up of the 1-degree reduction on single GPU, in Figure \ref{fig:h1} we compare the impact of 1-degree reduction on computation and communication
times on distributed systems. It is worth noting that with 4 GPUs the problem is computation-bound therefore the reduction of the total execution time is limited to the gain obtained on the computation. The improvement on the communication is more evident, for example, on the R-MAT graph with SCALE 20 EF 4, where the communication time with 1-degree on is halved with respect to the case with 1-degree turned off (see the second bar chart on Figure \ref{fig:h1}).
In Figure \ref{fig:h2} we show the performance of the 2-degree heuristics presented in Section \ref{sec:2degree} and in general the impact of heuristics in betweenness computation.
We focused on road networks since they present a significant number of 1-degree and 2-degree vertices.
In the y-axis, we report the number of vertices processed exploiting the techniques proposed in the present work.
For example, with no heuristics enabled, all the vertices of the graph
must be processed by MGBC (blue bar). On the other hand, the red and
transparent stacks represent the vertices processed by 1-degree and
2-degree heuristics without computing the BC explicitly.
The sum of the stacks must be equal to the total number of vertices of the graph (for RoadNet-PA $n=1090920$).
In the y2-axis, we report the total execution time (expressed in hours) for each heuristics. In particular
\begin{itemize}
 \item MGBC-H0 represents traditional MGBC without any heuristics turned on.
 \item MGBC-H1 exploits the 1-degree reduction.
 \item MGBC-H2 performs MGBC with 2-degree heuristics based on DMF techniques.
 \item MGBC-H3 combines 1-degree reduction and 2-degree heuristics.
\end{itemize}
The data reported are obtained running the experiments on Drake in
single GPU configuration. With H0, MGBC performs shortest paths counting, dependency accumulation and betweenness update procedure for each vertex in the graph \footnote{The disconnected vertices are also taken into account.}. For RoadNet-PA, the average time to perform these steps is $0.071$ seconds whereas the time to solution is about 21 hours.
With 1-degree turned on, $\sim 17\%$ of vertices are removed from the graph and their BC score contributions are directly computed from their neighbors.
We remark that  the procedure reduces both the number of vertices to traverse and the number of the vertices to perform MGBC.
MGBC-H1 is $17\%$ faster than MGBC-H0, in line with the percentage of 1-degree vertices. In this case, the improvement is
mainly due to the reduction in the number of MGBC execution.
On networks with a different topology, like the {\em com-youtube} graph, the improvement may be greater due also to a significant reduction in the total number of vertices to be visited.
Although the percentage of 2-degree vertices is $7\%$, we are able to handle only $5\%$ of them with a $5\%$ improvement in terms of MGBC performance (see H2 bar in Figure \ref{fig:h2}). The reason is that $2\%$ of 2-degree vertices share one or both neighbors. In this case, due to our implementation of DFM, we cannot augment the betweenness score of all 2-degree vertices.
As a matter of fact, on the contrary to 1-degree reduction, the 2-degree heuristics allows achieving a linear improvement depending only on the number of skipped Brandes' computations.
By combining H2 and H3 heuristics, we can achieve an improvement that is not just their sum, since the preprocessing of the 1-degree reduction increases the number of 2-degree vertices.
Basically 3-degree vertices which have a 1-degree neighbor become 2-degree after the 1-degree preprocessing step.
In our experiment we have $\sim 8\%$ of 2-degree vertices added.
Although the number of 1-degree vertices processed in H3 configuration are the same, the betweenness score of 2-degree vertices is twice ($10\%$) if compared to the H2 case.
The total number of vertices for which we avoid performing a round of MGBC is composed as follows: $17\%$ (due to 1-degree reduction) and $10\%$ computed by 2-degree heuristics.
By comparing with MGBC-H0, as expected the total improvement in terms of performance of MGBC-H3 is about $27\%$.

\begin{table*}[]\scriptsize
\begin{center}
\begin{tabular}{|c|c|c|c|}
\hline
\textbf{{\em fr}}        & \textbf{1}      & \textbf{64}     & \textbf{128}\\
\hline
Time (hours)& $211$  & $3.5$  & $1.8$   \\
GTEPS       & $0.94$ & $56.2$ & $111.60$ \\
\hline
\end{tabular}
\captionof{table}{Total time to compute exact BC for the Orkut graph with $fd =2$.}
\label{tab:cluster}
\end{center}
\end{table*}

\begin{table*}[htbp!]\scriptsize
\begin{center}
\resizebox{0.98\textwidth}{!} {
\begin{tabular}{|c|c|c|c|c|c|}
\hline
\textbf{Graph}  & \textbf{1-degree(\%)} & \textbf{Total time(hour)} & \textbf{Mean time(sec)} & \textbf{Preprocessing(sec)} & \textbf{Speed-up}\\
\hline
com-Youtube      & $53$  & $1.4$   ($3.9$)    & $0.0098$($0.012$) & $0.62$   & $2.8$x\\
     R-MAT EF4   & $13.6$& $1.1$   ($1.8$)    & $0.012$ ($0.015$)  & $0.312$ & $1.8$x\\
     R-MAT EF16  & $13.3$& $2.9$   ($4.1$)    & $0.021$ ($0.023$)  & $1.237$ & $1.4$x\\
     R-MAT EF32  & $12.1$& $5.0$   ($6.6$)    & $0.029$ ($0.032$)  & $2.449$ & $1.3$x\\
\hline
\end{tabular}
}
\caption{Impact on BC processing time due to 1-degree reduction. The value reported in parenthesis are referred to MGBC with 1-degree off.}
\label{tab:1degree}
\end{center}
\end{table*}

\begin{table*}[htbp!]\scriptsize
\begin{center}
\begin{tabular}{|c|c|c|c|c|c|}
\hline
\textbf{Graph}  & \textbf{Total time(hour)} & \textbf{Mean time(sec)} & \textbf{Traditional Execution} &\textbf{1-degree} &\textbf{2-degree}\\
\hline
MGBC-H0  & $21.8$ & $0.071$ & $1090920$ ($1090920$) & $0$ ($188317$)      & $0$ ($77265$)      \\
MGBC-H1  & $18.0$ & $0.070$ & $902603$  ($1090920$) &$188317$ ($188317$)  & $0$ ($77265$)     \\
MGBC-H2  & $20.8$ & $0.068$ & $1029219$ ($1090920$) & $0$ ($188317$)      & $61701$ ($77265$)  \\
MGBC-H3  & $15.9$ & $0.062$ & $791294$  ($1090920$) & $188317$ ($188317$) & $165788$ ($111309$)\\
\hline
\end{tabular}
\caption{Impact of heuristics on the exact Betweenness Computation on RoadNet-PA. The numbers in parenthesis represent the total number of vertices that may be computed by heuristics.}
\label{tab:h3}
\end{center}
\end{table*}

\section{Conclusions and future work}
\label{sec:conclusion}
We proposed a fast, communication-efficient, algorithm for the computation of betweenness centrality on Multi-GPU systems on unweighted graphs.
Our solution encapsulates three different levels of parallelism by
combining a fine- and coarse-grained approach using GPU accelerators.
In particular, sub-clustering allows reducing communication cost since the processors in the same sub-cluster are involved in the communication at the same time.
We also provide a technique to avoid exchanging predecessors during
traversal steps. This solution allows reducing the exchange of data from $\mathcal{O}(m)$ to $\mathcal{O}(n)$ regardless the partitioning strategy adopted.
Furthermore overlap optimization enables to speed-up the communication of sigma-delta among the GPUs.
The proposed algorithm has not only a single GPU performance comparable to state-of-the-art implementations, but it is able to scale up to 256 GPUs enabling the BC computation of large scale graphs, both real-world like Twitter or Friendster and R-MAT with scale up to 28.
We also provided an optimization to amortize the computation cost introduced by the thread-data mapping technique. This solution allows having a perfect load balancing among threads without paying extra computation costs in the dependency accumulation step.
We also investigated the impact of heuristics on betweenness centrality computation by providing comprehensive experiments.
In particular, on the contrary to previous works, we extended the 1-degree reduction heuristics on distributed systems and evaluated the impact on both computation and communication. Furthermore our solution supports the betweenness centrality computation on graphs with more connected components.
We presented a novel heuristics for 2-degree vertices based on an innovative algorithm (DFM) where the betweenness contributions are augmented from its two neighbors without performing the Brandes' algorithm explicitly.
We also provided a theoretical result which allows building a single source shortest path from a vertex if the shortest path trees of its own adjacencies are known.
Experimental results validated the effectiveness of our approach. The heuristics offers a speed-up that is, at least,  proportional to the number of skipped vertices.
Actually, a greater improvement can be obtained by combining 1-degree and 2-degree heuristics, since this allows deriving the BC score of particular 3-degree vertices as well.
\newline For the future,  we are investigating other heuristics. Moreover, we expect to release our code in the public domain to offer a tool able to compute BC on very large scale graphs.

\section*{Acknowledgment}
The authors would like to thank Mauro Bisson, Massimiliano Fatica, Andrea Formisano, Enrico Mastrostefano and Everett H. Phillips for very useful discussions and suggestions.
Finally, the authors would like to thank the Swiss National Supercomputing Centre for access and support to the ``Piz Daint'' cluster.



\end{document}